\documentclass[12pt,preprint]{aastex}
\shorttitle{$\gamma$-rays From FR I Jets }
\shortauthors{Stawarz, Sikora \$ Ostrowski}
\usepackage{natbib}

\begin{document}

\title{High Energy $\gamma$-rays From FR I Jets }

\author{\L . Stawarz}
\affil{Obserwatorium Astronomiczne, Uniwersytet Jagiello\'{n}ski, \\
ul. Orla 171, 30-244 Krak\'{o}w, Poland}
\email{stawarz@oa.uj.edu.pl}
\author{M. Sikora}
\affil{Centrum Astronomiczne im. M. Kopernika, \\
ul. Bartycka 18, 00-716 Warszawa, Poland\\
also at Stanford Linear Accelerator Center, Stanford, CA 94309-4349}
\author{and \\
M. Ostrowski}
\affil{Obserwatorium Astronomiczne, Uniwersytet Jagiello\'{n}ski, \\
ul. Orla 171, 30-244 Krak\'{o}w, Poland\\
also at Stanford Linear Accelerator Center, Stanford, CA 94309-4349}

\begin{abstract}
Thanks to {\it Hubble} and {\it Chandra} telescopes, some of the large scale
 jets in extragalactic radio sources are now being observed at optical and X-ray
 frequencies. For the FR I objects the synchrotron nature of this emission is
 surely established, although a lot of uncertainties -- connected for example
 with the particle acceleration processes involved -- remain. In this paper we
 study production of high energy $\gamma$-rays in FR I kiloparsec-scale jets by
 inverse-Compton emission of the synchrotron-emitting electrons. We consider
 different origin of seed photons contributing to the inverse-Compton scattering,
 including nuclear jet radiation as well as ambient, stellar and circumstellar
 emission of the host galaxies. We discuss how future detections or non-detections
 of the evaluated $\gamma$-ray fluxes can provide constraints on the unknown large
 scale jet parameters, i.e. the magnetic field intensity and the jet Doppler factor.
 For the nearby sources Centaurus A and M 87, we find measurable fluxes of TeV
 photons resulting from synchrotron self-Compton process and from comptonisation
 of the galactic photon fields, respectively. In the case of Centaurus A, we also
 find a relatively strong emission component due to comptonisation of the nuclear
 blazar photons, which could be easily observed by {\it GLAST} at energy $\sim 10$
 GeV, providing important test for the unification of FR I sources with BL Lac
 objects.
\end{abstract}


\section{Introduction}

$\gamma$-rays are detected from many galactic and extragalactic sources,
 including pulsars, supernova remnants, gamma ray bursts and blazars. It is
 believed that multiwavelength emission of the latest type of the mentioned
 objects is due to small scale highly relativistic jets, which are formed near
 active galactic nuclei of host galaxies and are inclined at small angles to the
 line of sight \citep[e.g.,][]{bre90}. High observed bolometric luminosities of
 blazars (typically $10^{44} - 10^{48}$ erg/s) are often dominated by
 $\gamma$-ray emission, what enabled to observe many of them by {\it EGRET} in
 $\varepsilon_{\gamma} \sim 0.1 - 10$ GeV energy range \citep{mon95,har99}. Some
 nearby low-luminosity blazars, like the best known Mrk 421 and Mrk 501, are
 also confirmed to be sources of very high energy (VHE) $\gamma$-rays with
 $\varepsilon_{\gamma} > 100$ GeV \citep[respectively]{pun92,qui96}. Detection
 of such VHE radiation by ground-based instruments was possible due to
 development of imaging atmospheric Cherenkov telescopes, {\it IACTs}
 \citep[see, e.g.,][]{aha98,vol03}.

The observed broad-band blazar emission allows one to determine physical
 parameters of nuclear AGN jets. $\gamma$-ray observations are crucial in this
 respect, providing important constraints on the involved particle acceleration
 processes, or on the characteristic spatial scales for the blazar phenomenon
 \citep[e.g.,][]{sik02,kin02}. However, observations at VHE range are
 complicated because propagation effects are important for the high energy
 photons created at cosmological distances \citep{nik62,gou66,ste92}, and also,
 obviously, because of a low photon statistics in $\gamma$-ray telescopes.
 Among the other issues, one should mention here a case of VHE emission from
 Mrk 501, which extends up to energies $\varepsilon_{\gamma}> 10$ TeV, which
 seems to be in conflict with some models of cosmic infrared background (CIB)
 radiation \citep[see, e.g.,][]{pro00, aha01}. Detailed studies of cosmic
 background photon fields and their interactions with $\gamma$-rays during their
 propagation, as well as future $\gamma$-ray missions, will possibly answer some
 of the questions connected with production of VHE radiation in blazars.

Contrary to blazar parsec-scale jets, their large scale counterparts extending
 from a few to a few hundreds of kiloparsecs from active galactic nuclei were
 usually studied only at radio frequencies \citep{bri84}. Recently, however,
 {\it Hubble} ({\it HST}) and {\it Chandra} telescopes gathered new information
 about optical and X-ray emission of some of these objects. For several hundreds
 of known radio jets, only about 20 are observed at optical
 frequencies\footnote{see homepage
 \texttt{http://home.fnal.gov/\~{}jester/optjets} by S. Jester and references
 therein}. Most of them are relatively short and faint, with only a few
 exceptions like the one in M 87 allowing for detailed spectroscopic and
 morphological studies \citep{mei96,per99,per01}. Surprisingly, the large scale
 jets can be also very prominent in X-rays. Up to now, more than 20 jets were
 detected by {\it Chandra} within the $1 - 10$ keV energy range\footnote{see
 homepage \texttt{http://hea-www.harvard.edu/XJETS} by D. Harris and references
 therein}, although the nature of this emission is still under debate
 \citep[e.g.,][]{tav00,bai01,cel01,aha02,car02,hk02,sta02,sah03,tav03}. In
 general, both optical and X-ray observations question in many aspects the
 standard models for the extragalactic jets emission.

As the large scale jets in radio galaxies are confirmed sources of optical and
 X-ray photons, one can ask if they can be also observed at $\gamma$-ray
 energies. Such observations could impose significant constraints on jet
 parameters (in analogy to the blazar sources), as well as for the unification
 scheme of radio-loud AGNs. In the present paper we make predictions regarding
 production of high energy $\gamma$-rays in FR I jets by the inverse-Compton
 scattering in a framework of one electron population model. In discussion, we
 refer to the objects observed by {\it Chandra}, as the X-ray observations of
 these recently widely studied sources impose important constraints on the jet
 synchrotron emission, crucial in determining the high energy photon flux. We
 find, that for the jet parameters inferred from radio-to-X-ray observations,
 only nearby Centaurus A and M 87 jets are expected to be observed at GeV --
 TeV energies by near future $\gamma$-ray missions (cf. discussion in section
 3). However, it is possible that more distant objects will be also detected,
 if the unconstrained jet parameters (like magnetic field and kinematic
 factors) differ from typically assumed values (sections 2.3.1 -- 2.3.3).

The already mentioned propagation effects can seriously affect the observed VHE
 spectra of distant objects, as interaction of VHE $\gamma$-rays with the CIB
 photon field is expected to result in their efficient absorption due to
 photon-photon pair creation. CIB radiation is created by infrared emission of
 forming stars in early cosmological epochs and the following reprocession by
 ambient dust. The origin, evolution and detailed energy distribution of CIB are
 still uncertain, mostly because of observational difficulties \citep[for a
 recent review see][]{hau01}. The existing controversies, particularly important
 in a TeV-absorption range (at $\nu_{CIB} \sim 10^{13}$ Hz) are however not
 crucial for our discussion, as we limit ourselves to relatively close object.
 For illustration, we refer to work by \citet[and references therein]{kne02} who,
 among the others, studied opacity of the Universe to $\gamma$-rays and evolution
 of the diffusive infrared-to-ultraviolet background radiation field.

Below, in section 2, based on optical and X-ray observations we reconstruct and
 comment `typical' electron energy distribution of FR I {\it Chandra} jets. We
 estimate different photon fields contributing to the inverse-Compton scattering
 of these electrons, and next evaluate and discuss the resulting fluxes and
 photon break energies in the Thomson regime. Section 3 contains detailed
 analysis of the $\gamma$-ray emission from the nearby sources Centaurus A and M
 87. Final conclusions are presented in section 4.

\section{Emission from the large scale jets in FR I sources}

\subsection{Synchrotron emission: radio-to-X-rays}

{\it Chandra} X-ray observatory detected about 10 jets displaying the FR I
 large scale morphology. They include the ones in radio galaxies M 87
 \citep{mar02,wil02}, Centaurus A \citep{kra00,kra02}, 3C 129 \citep{har02a},
 3C 31 \citep{har02}, PKS 0521-365 \citep{bir02}, 3C 270 \citep{chi03}, M 84
 \citep{har02b}, 3C 66B \citep{har01}, B2 0206+35 and B2 0755+37 \citep{wor01}.
 All of these objects are located relatively nearby, with a distance ranging
 from $3.4$ Mpc (Centaurus A) to more than $300$ Mpc (PKS 0521-315 at redshift
 $z = 0.055$). All of them share many spectral and morphological similarities
 briefly summarised below.

X-ray jets in FR I radio sources are usually quite short (projected length
 $\sim 1 - 4$ kpc). Typically, they are composed of diffusive knots with a
 spatial scale $R \sim 0.1$ kpc, although a strong inter-knot emission is also
 sometimes present. X-ray jet morphology corresponds \emph{roughly} to the radio
 morphology, and -- in cases of M 87, PKS 0521-315, 3C 66B, 3C 31 and B2 0755+37
 observed by {\it HST} -- to the optical one. A noted difference between these
 pictures are spatial offsets of some knot maxima as measured at X-rays and at
 radio/optical frequencies (up to $\sim 0.008$ kpc in M 87, $\sim 0.08$ kpc in
 Centaurus A and $\sim 0.2$ kpc in 3C 66B). Also, X-ray jets (knots) seem to be
 narrower than their radio/optical counterparts (M 87, Centaurus A). Except of
 the weakest and the smallest objects (3C 270 and M 84), the observed X-ray
 luminosities of the discussed jets are $L_X \sim 10^{39} - 10^{42}$ erg/s.

In most cases, multiwavelength observations of knot regions in FR I
 {\it Chandra} jets allow one to construct broad band radio-to-X-ray spectral
 energy distribution. All of the inferred knot spectra seem to be similar, and
 in addition difficult to be explained in a framework of standard jet models.
 A spectral index of the radio emission (defined in a way $S_{\nu} \propto
 \nu^{- \alpha}$, where $S_{\nu}$ is the energy flux spectral density) is
 always close to $\alpha_R \sim 0.6$. The optical emission obviously belongs
 to the synchrotron continuum, with the spectral index steepening between
 infrared and optical frequencies at about $\Delta \alpha \sim 0.7$ (3C 66B,
 PKS 0521-315, M 87). This agrees well with the X-ray spectral index measured
 by {\it Chandra}, which, for FR I jets, is on average $\alpha_X \sim 1.3$
 (ranging from $\sim 1.1$ in 3C 31 to $\geq 1.5$ in
 some knots of M 87 and Centaurus A). Thus, one can generalize, that in the
 discussed objects the radio-to-optical power-law slope is always less than
 unity ($\alpha_{RO} \sim 0.5 - 0.8$), while the optical-to-X-ray power-law
 slope is always much larger ($\alpha_{OX} \sim 1.1 - 1.5$). Such spectral
 behaviour seems to be universal for all FR I X-ray jets, with a spectral
 break ($br$) -- where most of the synchrotron ($syn$) power is emitted --
 placed usually around $\nu_{syn, \, br} \sim 10^{14}$ Hz or, eventually
 (M 87), at higher frequencies.

The steep X-ray spectral indices of FR I large scale jets exclude an
 inverse-Compton scattering as a mechanism responsible for the observed
 X-ray emission. Instead, radio-to-X-ray spectral properties indicate
 synchrotron origin of detected $1 - 10$ keV photons. If this is the case, than
 in the discussed objects unknown acceleration process operates, which results
 in characteristic and \emph{universal} for FR I jets electron energy
 distribution which can be approximated as a broken power-law:
\begin{equation}
n'_e (\gamma) \propto \left\{ \begin{array}{ccc} \gamma^{- p} & {\rm for} &
 \gamma_0 < \gamma < \gamma_{br} \\ \gamma^{- ( p \, + \, 2 \, \Delta \alpha)} &
 {\rm for} & \gamma_{br} < \gamma < \gamma_{max} \end{array} \right. \quad ,
\end{equation}
where $\gamma$ is the electron Lorentz factor\footnote{Hereafter we follow the
 notation with the primed quantities measured in the jet comoving frame and the
 bare ones if given in the observer rest frame (neglecting cosmological
 corrections). However, we do not prime the magnetic field induction $B$ as well
 as electron Lorentz factors $\gamma$, noting instead that they always refer to
 the emitting plasma rest frame.}, $\Delta \alpha$ is the spectral break in the
 synchrotron continuum and $p = 2 \, \alpha_R + 1 \sim 2.2$. Optical and X-ray
 observations play a crucial role in determining parameters of the above
 electron spectrum, i.e. its normalisation (for the assumed energy equipartition
 between electrons and the jet magnetic field) and the actual values of
 $\gamma_{br}$ and $\gamma_{max}$, which in turn determine the inverse-Compton
 $\gamma$-ray jet radiative output. For non-relativistic jet velocities, a
 typical $\nu_{syn, \, br} \sim 10^{14} - 10^{15}$ Hz and an equipartition
 magnetic field $B \sim 10^{-4}$ G, one gets $\gamma_{br} \sim 10^5 - 10^6$ and
 $\gamma_{max} > 10^7$. Discussion of the acceleration processes creating such
 high energy electrons in the FR I jets is beyond the scope of this paper. Let
 us note however, that it is possible for $n'_e(\gamma)$ to posses a more
 complicated form than the one given by equation 1, involving for example
 spectral pile-ups at the maximum electron energies due to efficient electron
 acceleration and rapid radiative cooling \citep{der02,sta02}. In either case,
 as the observed X-ray luminosity of the FR I kpc-scale jets is lower than the
 optical luminosity, most of the electrons energy is concentrated in the range
 near the electron break Lorentz factor $\gamma_{br}$.

It is interesting to compare $\nu_{syn, \, br}$ characteristic for the kpc-scale
 FR I jets with critical ($cr$) synchrotron frequency $\nu_{syn, \, cr}$
 resulting from the interplay between electron synchrotron cooling and dynamical
 evolution of the emitting region. The time scale for the former process can be
 estimated as $t'_{loss} \sim \gamma / | \dot{\gamma} |_{syn}$, where $m c^2 \,
 | \dot{\gamma} |_{syn} = {4 \over 3} \, c \, \sigma_T \, U'_B \, \gamma^2$ is a
 mean rate of electron energy losses due to synchrotron emission, and $U'_B =
 B^2 / 8 \pi$ denotes comoving jet magnetic field energy density. The second
 time scale is simply $t'_{dyn} = t_{dyn} / \Gamma \sim r / c \, \Gamma$, where
 $\Gamma = (1 - \beta^2)^{-1/2}$ is the jet bulk Lorentz factor, $\beta c$ is a
 jet bulk velocity and $r$ is the distance of the kpc-scale structure from the
 jet base at $r_0 \ll r$ from the active galactic nucleus. A critical electron
 Lorentz factor is defined by an equality $t'_{dyn} \approx t'_{loss}(\gamma =
 \gamma_{cr})$. Thus, assuming initial injection of the power-law electron
 energy distribution at $r_0$, one expects that at the distance $r$ the electron
 spectrum steepens for $\gamma > \gamma_{cr}$ because of the efficient radiative
 cooling of these electrons. The observed critical synchrotron frequency is then
 $\nu_{syn, \, cr} = \nu'_{syn, \, cr} \, \delta \propto \gamma_{cr}^2 \, B \,
 \delta$, where
\begin{equation}
\delta = {1 \over \Gamma \, (1 - \beta \, \cos \theta)}
\end{equation}
is the jet Doppler factor and $\theta$ is the jet inclination angle to the line
 of sight. Assuming that the jet magnetic field scales with the distance as $B =
 B_0 \, (r_0 / r)$, what conserves the Poynting flux $L'_B \propto R^2 \, U'_B$
 in the expanding jet with an opening angle $\varphi \sim const$ and a radius $R
 \sim \varphi \, r$, one obtains
\begin{equation}
\nu_{cr} \sim 4 \cdot 10^{14} \, r_1 \, \delta \, \Gamma^2 \quad {\rm Hz} \quad
 ,
\end{equation}
where we put $r_0 = 1$ pc, $B_0 = 0.1$ G and $r_1 \equiv r / 1 \, {\rm kpc}$.
 Note, that with the above values of $B_0$ and $r_0$ characteristic for the
 blazar sources, the kpc-scale jet magnetic field is expected to be $B \sim
 10^{-4}$ G, consistently with the equipartition value. Hence, in a case of
 a non-relativistic kpc-scale jet velocity, $\nu_{syn, \, cr} \sim \nu_{syn,
 \, br} \sim 10^{14}$ Hz. However, the spectral break resulting from the
 considered process is $\Delta \alpha_{cr} = 0.5$ \citep{kar62}, i.e. slightly
 less than the one required for the FR I {\it Chandra} jets. Possibly, the
 observed steep spectral break is a signature of the electron synchrotron
 cooling in the spatially \emph{inhomogeneous} magnetic field
 \citep{cav80,col88}. Note also, that in the cases of relativistic jet bulk
 velocities, the observed value of $\nu_{syn, \, cr}$ can be significantly
 higher than $10^{14}$ Hz (cf. the case of M 87, section 3.2).

Radio observations of FR I jets indicate moderate or even weak beaming
 \citep[e.g.,][]{lai99}. However, one cannot exclude possibility that the bulk
 Lorentz factors of kpc-scale jets in weak radio galaxies are of order of a few.
 This idea is supported by {\it HST} observations of superluminal motions and
 {\it Chandra} detection of significant X-ray variability at kpc-scales in M87
 \citep[respectively]{bir99,har03}. In fact, a small number of detected optical
 and X-ray jets in radio galaxies as compared to the number of known radio jets
 suggests that relativistic beaming effects can play an important role in these
 sources \citep{spa95,sca02,jes03}. Including a relativistic correction, the
 electron break Lorentz factor is
\begin{equation}
\gamma_{br} \sim 5 \cdot 10^5 \, \left( { \nu_{syn, \, 14} \over B_{-4} \,
 \delta} \right)^{1/2} \quad ,
\end{equation}
where $\nu_{syn, \, 14} \equiv \nu_{syn, \, br} / 10^{14}$ Hz and $B_{-4} \equiv
 B / 10^{-4}$ G.

\subsection{Radiation fields within the kpc-scale jet}

Let us consider a kpc-scale ($r \sim 1 - 4$ kpc) relativistic jet with the bulk
 Lorentz factor $\Gamma$, inclined at an angle $\theta$ to the line of sight.
 The energy density of the jet magnetic field, as measured in the emitting
 region rest frame, is
\begin{equation}
U'_B = {B^2 \over 8 \, \pi} \sim 4 \cdot 10^{-10} \, B^2_{-4} \quad {\rm erg \,
 cm^{-3}} \quad .
\end{equation}
As mentioned above, the equipartition value inferred from the synchrotron
 emission of FR I knot regions is typically $B_{eq} \sim 10^{-4}$ G, and it
 corresponds to the observed synchrotron jet luminosities $L_{syn} \sim 10^{40}
 - 10^{42}$ erg/s. The observed luminosity is related to the total emitted
 synchrotron power $L'_{syn}$ by the relation
\begin{equation}
L_{syn} = g_{cj / mb}(\Gamma, \theta) \, L'_{syn} \quad ,
\end{equation}
where
\begin{equation}
g_{cj / mb}(\Gamma, \theta) = \left\{ \begin{array}{ccc} \delta^3 / \Gamma &
 (cj) \\ \delta^4 & (mb) \end{array} \right.
\end{equation}
in cases of a continuous jet ($cj$) or a moving single radiating blob ($mb$),
 respectively \citep[see also Appendix A]{sik97}\footnote{Note, that as
 discussed in Appendix A, in both cases the equipartition magnetic field
 measured in the emitting region rest frame is related to the equipartition
 value computed for no beaming by the relation $B_{eq} = B_{eq, \, \delta = 1}
 \, \delta^{-5/7}$.}. Below we analyse these both possibilities as limiting
 models for the kpc-scale knots. We also consider synchrotron luminosity at a
 given break frequency $\nu_{syn, \, br}$, hereafter denoted as $[\nu
 L_{\nu}]_{syn, \, br}$, rather than the bolometric one $L_{syn}$, with
 transformations between $[\nu L_{\nu}]_{syn, \, br}$ and $[\nu'
 L'_{\nu'}]_{syn, \, br}$ the same as given in equations 6 - 7. For the broken
 power-law synchrotron spectrum assumed here, with $\alpha \sim 0.6$ and $\Delta
 \alpha \sim 0.7$ (cf. equation 1) one has a bolometric correction $L_{syn} /
 [\nu L_{\nu}]_{syn, \, br} \approx 6$.

In addition to the synchrotron emission extending to the X-rays, the
 FR I jets produce also the high energy radiation by inverse-Compton scattering
 of ambient photon fields, like, for example, the cosmic microwave background
 (CMB) radiation or the synchrotron emission produced by the jet itself.
 At the kiloparsec distances from the galactic center another
 important sources of ambient photons are the active galactic nucleus and the
 host galaxy. The AGN radiation, produced at small distances from the central
 region, illuminates the large scale jet almost exactly from behind. This
 radiation consists of isotropically emitted component connected with thermal
 gas and/or dust heated by the central source (the narrow line region and the
 dusty nuclear torus) plus an anisotropic blazar-like emission due to small
 scale highly relativistic nuclear jet. The narrow line emission and the
 radiation of dusty nuclear tours, typically weak or even absent in FR I radio
 galaxies \citep[e.g.,][]{chi99}, are always negligible at distances $r \geq 1$
 kpc from the active nucleus \citep[section 2.2.2, see also][]{cel01}.
 Instead, as discussed below, comptonisation of starlight and extended dust
 emission of the host galaxy and, in some cases, comptonisation of the `hidden'
 blazar radiation can dominate $\gamma$-ray output of the FR I large scale jets.

\subsubsection{Synchrotron photons}

For a simple evaluation of the synchrotron photons energy density in the jet
 comoving frame, we consider an approximately cylindrical knot region, with a
 radius $R \leq 0.1$ kpc and a deprojected observed length $l \sim R$. In the
 case of a continuous jet one has $l' = l \, \Gamma > R$, and therefore
\begin{equation}
U'_{syn} = {[\nu' L'_{\nu'}]_{syn, \, br} \over 2 \pi \, R \, l' \, c} = {[\nu
 L_{\nu}]_{syn, \, br} \over 2 \pi \, R \, l \, c} \, {1 \over \delta^3} \sim
 5.6 \cdot 10^{-11} \, {[\nu L_{\nu}]_{syn, \, 42} \over R_{-1}^2} \, {1 \over
 \delta^3} \quad {\rm erg \, cm^{-3}} \quad ,
\end{equation}
where $[\nu L_{\nu}]_{syn, \, 42} \equiv [\nu L_{\nu}]_{syn, \, br} / 10^{42} \,
 {\rm erg \, s^{-1}}$ and $R_{-1} \equiv R / 0.1 \, {\rm kpc}$. If the emitting
 kpc-scale jet region is in fact a moving source, then $l' = l \, / \, \delta =
 R \, / \, \delta$, and the jet comoving energy density of the synchrotron
 emission is
\begin{equation}
U'_{syn} = {[\nu' L'_{\nu'}]_{syn, \, br} \over 2 \pi \, R^2 \, c} = {[\nu
 L_{\nu}]_{syn, \, br} \over 2 \pi \, R^2 \, c} \, {1 \over \delta^4} \sim 5.6
 \cdot 10^{-11} \, {[\nu L_{\nu}]_{syn, \, 42} \over R_{-1}^2} \, {1 \over
 \delta^4} \quad {\rm erg \, cm^{-3}} \quad .
\end{equation}
The jet comoving synchrotron break frequency is $\nu'_{syn, \, br} = \nu_{syn,
 \, br} / \delta$. Note, that for the approximately constant $\varphi \sim 0.1$,
 the jet radius at a distance $r \sim 1$ kpc from its base is expected to be $R
 \sim 0.1$ kpc, consistently with the value considered here.

\subsubsection{Radiation of the active nucleus}

At first, let us consider the blazar-like ($bl$) emission of FR I active nuclei,
 which -- accordingly to the unification scheme \citep[e.g.,][]{urr95} -- belong
 to the low-luminous (BL Lac) blazar subclass. Due to relativistic effects, this
 radiation is strongly beamed into the cone with an opening angle $\theta_{bl}
 \sim 1 / \Gamma_{bl}$, where $\Gamma_{bl}$ is the nuclear jet bulk Lorentz
 factor. In order to evaluate energy density of this emission in the kpc-scale
 jet rest frame, we assume that it `enters' into the considered emitting region
 directly from the jet base. In other words, we neglect any possible
 misalignment between the pc-scale and the kpc-scale jets. The respective energy
 density of the blazar photons is then
\begin{equation}
U'_{bl} \sim {L_{bl}(0) \over 4 \pi \, r^2 \, c} \, {1 \over (2 \Gamma)^2 }
 \quad ,
\end{equation}
where $L_{bl}(0)$ is the isotropic luminosity of the blazar emission evaluated
 by the stationary observer located at $\theta = 0$ (Appendix B). With the
 spatial scales of the kpc-scale jet emitting region much larger than the
 characteristic nuclear scales ($R, \, r \gg r_0$), the discussed source of the
 seed photons for the inverse-Compton scattering should be considered as a {\it
 stationary} one. A typical short time scale of blazar variability justifies
 this statement. Therefore,
\begin{equation}
L_{bl}(0) = L_{bl} \, {\delta^3_{bl, \, \theta = 0} / \Gamma_{bl} \over
 \delta^3_{bl} / \Gamma_{bl}} \sim L_{bl} \, \left( {2 \Gamma_{bl} \over
 \delta_{bl}} \right)^3 \quad ,
\end{equation}
where $\delta_{bl} = [\Gamma_{bl} \, (1 - \beta_{bl} \, \cos \theta)]^{-1}$ and
 $L_{bl}$ is the isotropic luminosity of the considered stationary blazar source
 measured by the observer located at the angle $\theta$ to the jet axis. Hence,
\begin{equation}
U'_{bl} \sim {L_{bl} \, \Gamma_{bl}^3 \over 2 \pi \, r^2 \, c \, \delta_{bl}^3}
 \, {1 \over \Gamma^2} \sim 5.6 \cdot 10^{-13} \, {L_{bl, 42} \, \Gamma_{bl}^3
 \over r_{1}^2 \, \delta_{bl}^3} \, {1 \over \Gamma^2} \quad {\rm erg \,
 cm^{-3}} \quad ,
\end{equation}
where we put $L_{bl, \, 42} \equiv L_{bl} / 10^{42} \, {\rm erg \, s^{-1}}$
 \citep[cf. equation 1 in][]{cel01}.

Typically, for the high energy peaked BL Lacs (HBLs) the observed synchrotron
 luminosity is $10^{44} - 10^{46} \, {\rm erg \, s^{-1}}$, while for the low
 energy peaked BL Lacs (LBLs) it is in the range $10^{45} - 10^{47} \, {\rm erg
 \, s^{-1}}$ \citep{fos98}. An another subclass of blazar sources -- the flat
 spectrum radio quasars (FSRQs) -- have even higher observed luminosities.
 However, in a framework of the unification scheme FSRQs correspond to the FR II
 sources. The critical observed synchrotron frequency of blazar emission, which
 anticorrelates with the blazar synchrotron luminosity, is usually $10^{15} -
 10^{17}$ Hz for HBLs and $10^{13} - 10^{15}$ Hz for LBLs
 \citep[again][]{fos98}. Below, we take $L_{bl} \sim [\nu L_{\nu}]_{bl, \, br}$,
 where the blazar synchrotron break frequency in the jet comoving frame is
 $\nu'_{bl, \, br} \sim \nu_{bl, \, br}(0) / \Gamma$, and $\nu_{bl, \, br}(0) =
 \nu_{bl, \, br} \, (2 \Gamma_{bl} / \delta_{bl})$. We also assume, that the
 blazar source illuminates uniformly the whole kpc-scale jet emitting region.
 Note, that for the typical $\Gamma_{bl} \sim 10$, the half-angle of the blazar
 emission cone is equal to the assumed jet opening angle, $\theta_{bl} \sim
 \varphi \sim 0.1$.

As mentioned previously, the isotropic component of the nuclear ($nucl$)
 emission, i.e. radiation of narrow line region or dusty nuclear torus, are weak
 or even absent in FR I radio galaxies (isotropic luminosity $L_{nucl} <
 10^{42}$ erg/s). For the observer located at the distance $r$ from the active
 center and moving with the Lorentz factor $\Gamma$ along the jet axis, energy
 density of the considered circumnuclear isotropic emission scales in the same
 manner as the energy density of the blazar radiation -- in both cases the
 appropriate relations are $U'_{nucl} \sim L_{nucl} / 4 \pi r^2 c (2 \Gamma)^2$
 and the one given in equation 10, respectively. Therefore, as $L_{nucl} \ll
 L_{bl}(0)$, the isotropicly emitted nuclear photon fields are negligible as
 compared to the blazar emission.

\subsubsection{CMB and galactic photon fields}

AGNs are hosted by elliptical galaxies, whose extended optical and infrared
 emission contribute significantly to the total radiation intensity at the
 kiloparsec distances from the galactic centers. 

Ellipticals are known to contain $\sim 1 \, M_{\odot}$ stars evolving from the
 main sequence through the red-giant phase, hot interstellar gas heated to
 temperatures $\sim 10^7$ K and radiating at X-rays, and, finally, dust
 prominent at infrared frequencies. Photospheric emission from cool giant stars,
 which are concentrated within a few hundreds of parsecs from the galactic
 center (i.e. within a `central bulge' or a `galactic core'), dominates
 bolometric luminosities of the discussed galaxies and peaks in the range $\sim
 1 - 3 \, {\mu} \, {\rm m}$ \citep[e.g.,][]{kna92}. Total bolometric
 luminosities of giant ellipticals can be as high as $\sim 3 \cdot 10^{11} \,
 L_{\odot} \sim 10^{45}$ erg/s. Emission at wavelengths $5 - 20 \, \mu {\rm m}$
 is analogously distributed as the starlight and is most likely connected with
 dusty winds accompanying the red-giant stars and forming their dusty
 circumstellar envelopes \citep{kna92,tsa95}. Radio galaxies are in general more
 luminous at infrared wavelengths as compared to the normal ellipticals (by a
 factor of 2 -- 3 relative to the starlight emission), with $\sim 10 \, \mu {\rm
 m}$ luminosity being on average two orders of magnitude lower than the
 bolometric one \citep{kna90}. In addition to the circumstellar dust emission,
 ellipticals exhibit also an excess at far-infrared frequencies, peaking around
 $60 - 100 \, {\mu} \, {\rm m}$. If the cold dust responsible for production of
 these photons is optically thin, it must occupy regions that extend far from
 the galactic core (a few times the core radius), in order to avoid overheating
 by the starlight \citep{tsa96}. An alternative explanation can be provided by
 concentration of the cold dust in disks and lanes self-shielded from the
 starlight and located closer to the galactic center. The far-infrared excess is
 especially prominent in radio galaxies \citep{gol88,kna90}. However, as the
 origin and spatial distribution of this radiation is unknown, and also as the
 far-infrared emission shows significant variations from galaxy to galaxy, we
 do not discuss its contribution to the kpc-scale jet inverse-Compton emission.

In Appendix C, following \citet{tsa95}, we estimate the energy
 density of the starlight ($star$) photons to be $U_{star} \sim 10^{-9}$
 erg/cm$^3$ at the distance $\sim 1$ kpc from the center of the typical giant
 elliptical galaxy. Assuming for simplicity an approximately isotropic
 distribution of the stellar emission at this scale, in the jet comoving frame
 one has
\begin{equation}
U'_{star} \sim U_{star} \, \Gamma^2 \sim 10^{-9} \, \Gamma^2 \quad {\rm erg \,
 cm^{-3}}
\end{equation}
(Appendix B), with a characteristic starlight frequency $\nu'_{star} \sim
 10^{14} \, \Gamma$ Hz. The galactic dust emission ($dust$) is distributed
 analogously to the stellar radiation and hence its energy density in the jet
 rest frame is also amplified approximately by a factor $\Gamma^2$. In this
 paper we take
\begin{equation}
U'_{dust} \sim 0.01 \cdot U'_{star} \sim 10^{-11} \, \Gamma^2 \quad {\rm erg \,
 cm^{-3}}
\end{equation}
at $\nu'_{dust} \sim 3 \cdot 10^{13} \, \Gamma$ Hz. Finally, for redshifts $z
 \ll 1$ an analogous energy density of the blackbody CMB is equal to
\begin{equation}
U'_{CMB} = a \, T_{CMB}^4 \Gamma^2 \sim 4 \cdot 10^{-13} \, \Gamma^2 \quad {\rm
 erg \, cm^{-3}} \quad ,
\end{equation}
where $a = 7.53 \cdot 10^{-15}$ cgs and the observed CMB temperature is $T_{CMB}
 = 2.7$ K. The appropriate characteristic CMB photon frequency is $\nu'_{CMB}
 \sim 2 \cdot 10^{11}\, \Gamma$ Hz.

\subsection{Inverse-Compton emission: $\gamma$-rays}

With the evaluated photon fields in the emitting region rest frame, $U'_{seed}$,
 one can estimate the observed break luminosity of the appropriate
 inverse-Compton ($ic(seed)$) emission in the Thomson regime as
\begin{equation}
[\nu L_{\nu}]_{ic(seed), \, br} \sim f_{\pm / iso}(\Gamma, \theta) \, {U'_{seed}
 \over U'_B} \, [\nu L_{\nu}]_{syn, \, br} \quad .
\end{equation}
Presence of an additional factor $f_{\pm / iso}(\Gamma, \theta)$ is connected
 with the possible anisotropy of the external radiation fields within the jet
 comoving frame, in cases of its relativistic bulk velocities and/or
 non-isotropic photon distribution in the galaxy rest frame. As shown in
 Appendix D,
\begin{equation}
f_{\pm / iso}(\Gamma, \theta) = \left\{ \begin{array}{ccc} {3 \over 4} \, \left(
 \delta / \Gamma \right)^2 \, \left[ (1 + \mu) / (1+\beta) \right]^2 & (+) \\ {3
 \over 4} \, \left( \delta \, \Gamma \right)^2 \, \left[(1 - \mu) \, (1 + \beta)
 \right]^2 & (-) \\ 1 & (iso) \end{array} \right.
\end{equation}
respectively for the cases when the seed photons are distributed isotropically
 in the galaxy rest frame near the kpc-scale jet ($+$), when they illuminate the
 jet exactly from behind ($-$), or when they are isotropic \emph{in} the jet
 comoving frame ($iso$). Luminosity $[\nu L_{\nu}]_{ic(seed), \, br}$ is related
 to the observed flux of the inverse-Compton emission, $[\nu S_{\nu}]_{ic(seed),
 \, br}$, by the relation
\begin{equation}
[\nu S_{\nu}]_{ic(seed), \, br} = { [\nu L_{\nu}]_{ic(seed), \, br} \over 4 \pi
 \, D^2} \quad ,
\end{equation}
where $D$ is a distance to the source. The appropriate observed inverse-Compton
 break frequency due to $\gamma_{br}$ electrons scattering the seed photons with
 the observed characteristic (break) frequency
 $\nu_{seed}$ is
\begin{equation}
\nu_{ic(seed), \, br} = h_{\pm / iso}(\Gamma, \theta) \, \gamma_{br}^2 \,
 \nu_{seed} \quad ,
\end{equation}
where
\begin{equation}
h_{\pm / iso}(\Gamma, \theta) = \left\{ \begin{array}{ccc} \delta^2 \, (1 + \mu)
 / (1 + \beta) & (+) \\ \delta^2 \, (1 - \mu) \, (1 + \beta) & (-) \\ 4/3 &
 (iso) \end{array} \right.
\end{equation}
(Appendix E). Assuming that the scattering proceeds in the Thomson regime, 
\begin{equation}
\gamma_{br} \, \nu'_{seed} < {m c^2 \over h} \quad ,
\end{equation}
the inverse-Compton spectrum for $\nu_{ic(seed)} < \nu_{ic(seed), \, br}$ is a
 power-law with a spectral index corresponding to the radio-to-optical
 synchrotron continuum, $\alpha_{\gamma} \sim (p - 1)/2$ (equation 1), while for
 $\nu_{ic(seed)} > \nu_{ic(seed), \, br}$ it is expected to steepen by $\Delta
 \alpha$. The high energy cut-off in the observed $\gamma$-ray emission can
 result from the cut-off in the electron energy distribution ($\gamma_{max}$),
 from entering the Klein-Nishina (KN) regime or due to external absorption on
 the CIB radiation.

Below, we evaluate fluxes and break energies of the inverse-Compton emission on
 photon fields described in the previous section. At first we discuss the
 synchrotron self-Compton radiation, i.e. the inverse-Compton scattering of the
 synchrotron photons produced by the same electron population ($ic = ssc$; $seed
 = syn$). Next, we analyse the external-Compton ($ic \equiv ec$) process
 starting from comptonisation of the `hidden' blazar emission ($seed = bl$).
 Then we discuss the scattering of CMB photons and different radiation fields of
 the host galaxy ($seed = star, \, dust, \, CMB$).

\subsubsection{Synchrotron self-Compton emission}

From the equations 8 - 9 and 16 - 18 with $f_{iso}(\Gamma, \, \theta) = 1$, one
 obtains the observed SSC energy flux for the two considered cases of a
 continuous jet and a moving blob
\begin{equation}
[\nu S_{\nu}]_{ssc, \, br} \sim 1.2 \cdot 10^{-11} \, {[\nu L_{\nu}]_{syn, \,
 42}^2 \over B_{-4}^{2} \, R_{-1}^{2} \, D_{10}^{2}} \left\{ \begin{array}{ccc}
 \delta^{-3} & (cj) \\ \delta^{-4} & (mb) \end{array} \right. \quad {\rm {erg
 \over s \, cm^{2}}}
\end{equation}
respectively, where $D_{10} \equiv D / 10$ Mpc. Equations 19 - 20 with
 $h_{iso}(\Gamma, \, \theta) = 1$ give the observed SSC break energy
\begin{equation}
\varepsilon_{ssc, \, br} \sim 1.5 \cdot 10^{11} \, {\nu_{syn, \, 14}^2 \over
 B_{-4}} \, {1 \over \delta} \quad {\rm eV} \quad .
\end{equation}
The Thomson regime condition for the discussed process reads as
\begin{equation}
\nu_{syn, \, 14} < 2 \, B_{-4}^{1/3} \, \delta \quad .
\end{equation}

The SSC emission is determined by the observed parameters $[\nu L_{\nu}]_{syn,
 \, br}$ and $\nu_{syn, \, br}$, plus the emitting region linear size $R$, the
 (unknown) magnetic field $B$ and the kinematic factor $\delta$. Typically, for
 the FR I {\it Chandra} jets one has $[\nu L_{\nu}]_{syn, \, 42} < 1$, $R_{-1}
 \sim 1$ and the equipartition value (as computed for nonrelativistic bulk
 velocities) $B_{-4} \sim 1$. Hence, for $\nu_{syn, \, 14} \sim 1$ and the
 usually considered $\delta \sim 1$, one can put an upper limit $[\nu
 S_{\nu}]_{ssc, \, br} < 10^{-11} \, D_{10}^{-2} \, {\rm erg / s \, cm^2}$ at
 the critical photon energy $\varepsilon_{ssc, \, br} \sim 0.1$ TeV. Future {\it
 IACT} systems will be able to eventually detect such emission\footnote{with
 the detector sensitivity $\geq 10^{-13} \, {\rm erg / s \, cm^2}$ for $5 \sigma$
 detection threshold and 100 h observation \citep{aha98}} from the sources at
 the distances $D < 100$ Mpc. Higher SSC flux than discussed above can be
 produced for a given $[\nu L_{\nu}]_{syn, \, br}$ in the case of a strong
 departure from the equipartition, $B \ll B_{eq}$, and/or for a de-beaming of
 the jet emission, $\delta \ll 1$. However, for decreasing $B$ and $\delta$
 the effects connected with the KN regime are expected to become important
 (unless $\nu_{syn, \, 14} \ll 1$, equation 24), decreasing the efficiency of
 the SSC process.

\subsubsection{Comptonisation of the hidden blazar emission}

From the equations 12 and 16 - 20, one obtains the observed EC(bl) energy flux
\begin{equation}
[\nu S_{\nu}]_{ec(bl), \, br} \sim 1.2 \cdot 10^{-13} \, {[\nu L_{\nu}]_{bl, \,
 42} \, [\nu L_{\nu}]_{syn, \, 42} \, \Gamma_{bl}^3 \over B_{-4}^{2} \,
 r_{1}^{2} \, D_{10}^{2} \, \delta_{bl}^3} \, {f_{-} \over \Gamma^2} \quad {\rm
 {erg \over s \, cm^{2}}} \quad ,
\end{equation}
and the observed EC(bl) break energy
\begin{equation}
\varepsilon_{ec(bl), \, br} \sim 2.2 \cdot 10^{11} \, {\nu_{bl, \, 14} \,
 \nu_{syn, \, 14} \, \Gamma_{bl} \over B_{-4} \, \delta_{bl}} \, {h_{-} \over
 \delta} \quad {\rm eV} \quad .
\end{equation}
The Thomson regime condition for the considered process reads as
\begin{equation}
\Gamma_{bl} \, \delta_{bl}^{-1} \, \nu_{bl, \, 14} < \Gamma \, \delta^{1/2} \,
 B_{-4}^{1/2} \, \nu_{syn, \, 14}^{-1/2} \quad .
\end{equation}

The EC(bl) emission is determined by the jet parameters $[\nu L_{\nu}]_{syn, \,
 br}$ and $\nu_{syn, \, br}$ plus the distance from the blazar source $r$, the
 unknown magnetic field $B$, kinematic factors $\Gamma$ and $\theta$ and,
 finally, by properties of the blazar core $[\nu L_{\nu}]_{bl, \, br}$,
 $\nu_{bl, \, br}$ and $\Gamma_{bl}$. For the usually discussed $\delta \sim 1$
 and $\theta \geq 45^0$, typical $\nu_{syn, \, 14} \sim 1$ and the equipartition
 value $B_{-4} \sim 1$, one expects the observed EC(bl) break energy to be
 $\varepsilon_{ec(HBL), \, br} \sim 1 - 100$ TeV in the case of comptonisation
 of HBL radiation ($bl \equiv HBL$), and $\varepsilon_{ec(LBL), \, br} \sim 0.01
 - 1$ TeV in the case of LBL-like core emission ($bl \equiv LBL$). As the
 synchrotron break frequency of the blazar emission is anticorrelated with the
 blazar synchrotron luminosity (see section 2.2.2), one can put upper limits on
 the observed EC(bl) fluxes $[\nu S_{\nu}]_{ec(HBL), \, br} < 10^{-11} \,
 D_{10}^{-2} \, {\rm erg / s \, cm^2}$ for $\varepsilon_{ec(HBL), \, br} \sim 1$
 TeV (taking a rough estimate $L_{HBL, \, max}(0) \leq 10^{46}$ erg/s), and
 $[\nu S_{\nu}]_{ec(LBL), \, br} < 10^{-10} \, D_{10}^{-2} \, {\rm erg / s \,
 cm^2}$ for $\varepsilon_{ec(LBL), \, br} \sim 10$ GeV (with $L_{LBL, \, max}(0)
 \leq 10^{47}$ erg/s). In addition, the EC(HBL) emission is expected to be
 significantly decreased due to the KN effects, contrary to the EC(LBL)
 radiation. For the discussed standard jet parameters the latter one could be
 observed by {\it GLAST}\footnote{accordingly to the $10$ GeV $\it GLAST$
 sensitivity $\sim 10^{-12} \, {\rm erg / s \, cm^2}$ for $5 \sigma$ detection
 threshold in 1 year all-sky survey \citep{aha98}} from sources with $D < 100$
 Mpc. Similarly to the case of the SSC process, for a given $[\nu L_{\nu}]_{syn,
 \, br}$ the subequipartition magnetic field and/or large jet inclination
 increase the expected value of $[\nu S_{\nu}]_{ec(bl), \, br}$.

\subsubsection{Comptonisation of the galactic photon fields and CMB radiation}

From the equations 13 and 16 - 20, one obtains the observed EC(star) energy flux

\begin{equation}
[\nu S_{\nu}]_{ec(star), \, br} \sim 2.1 \cdot 10^{-10} \, {[\nu L_{\nu}]_{syn,
 \, 42} \over B_{-4}^{2} \, D_{10}^{2}} \, {f_{+} \, \Gamma^2} \quad {\rm {erg
 \over s \, cm^{2}}} \quad ,
\end{equation}
and the observed EC(star) break energy
\begin{equation}
\varepsilon_{ec(star), \, br} \sim 1.1 \cdot 10^{11} \, {\nu_{syn, \, 14} \over
 B_{-4}} \, {h_{+} \over \delta} \quad {\rm eV} \quad .
\end{equation}
The considered process proceeds within the Thomson regime for
\begin{equation}
\nu_{syn, \, 14} < 6 \, B_{-4} \, {\delta \over \Gamma^2} \quad .
\end{equation}

The EC(star) radiation is determined by the jet parameters $[\nu L_{\nu}]_{syn,
 \, br}$, $\nu_{syn, \, br}$, $B$ and $\delta$, as well as by properties of the
 host galaxy ($L_{star}$, etc). For non-relativistic jets, typical $\nu_{syn, \,
 14} \sim 1$, $[\nu L_{\nu}]_{syn, \, 42} < 1$ and $B_{-4} \sim 1$, the break
 energy of the considered emission is $\varepsilon_{ec(star), \, br} \sim 0.1$
 TeV, and the observed energy flux is $[\nu S_{\nu}]_{ec(star), \, br} <
 10^{-10} \, D_{10}^{-2} \, {\rm erg / s \, cm^2}$. However, this relatively
 strong radiation can be significantly decreased in the case of relativistic jet
 velocities and large jet inclinations to the line of sight due to both the KN
 and the Doppler effects (equations 30 and 28, respectively). The eventuall
 Doppler-hide results from the beaming pattern of the EC(star)
 emission, which -- contrary to the discussed before SSC and EC(bl) processes --
 is maximised for small jet viewing angles (for a given $[\nu L_{\nu}]_{syn, \,
 br}$ one has $[\nu S_{\nu}]_{ec(star), \, br} \propto f_{+} \, \Gamma^2 \sim
 \delta^2$; equation 28). Comptonisation of the galactic infrared emission, as
 well as of the CMB radiation, is not expected to suffer such a decrease
 connected with the KN regime. For example, the Thomson regime condition for the
 EC(dust) process can be rewritten as $\nu_{syn, \, 14} < 60 \, B_{-4} \, \delta
 \, \Gamma^{-2}$. For the standard nonrelativistic jet parameters, one can
 therefore put an upper limit $[\nu S_{\nu}]_{ec(dust), \, br} < 10^{-12} \,
 D_{10}^{-2} \, {\rm erg / s \, cm^2}$ at $\varepsilon_{ec(dust), \, br} \geq
 10$ GeV. At even lower photon energies the EC(CMB) emission dominates,
 proceeding well in the Thomson regime, with the maximum placed near
 $\varepsilon_{ec(CMB), \, br} \sim 0.1$ GeV. The observed flux of this
 radiation is relatively low, $[\nu S_{\nu}]_{ec(CMB), \, br} < 10^{-13} \,
 D_{10}^{-2} \, {\rm erg / s \, cm^2}$. Detection of this emission from the FR I
 jets located further than a few Mpcs (i.e. further than Centaurus A) by future
 $\gamma$-ray telescopes is therefore possible only if the jet Doppler factor is
 significantly larger than unity, and/or if $B \ll B_{eq}$. On the other hand,
 large values of $\delta$ are unlikely to occur in the FR I jets, although, as
 mentioned in section 2.1, the optical and X-ray jets can constitute the
 exceptions from this picture.

\section{Centaurus A and M 87}

The resulting inverse-Compton fluxes and break energies evaluated in the
 previous section depend on several free parameters. Because of relatively
 wide range of values covered by them, it is difficult to discuss
 $\gamma$-ray emission of the considered objects in general. Instead, a detailed
 analysis can be performed for each individual source. In this section, we
 discuss the large scale jet $\gamma$-ray radiative outputs of Centaurus A and M
 87. One may note at this point, that both sources are quite unique. They are
 relatively nearby, what allowed in the past to study their jets at different
 frequencies and scales, with high spatial and spectral resolutions. The large
 scale jet in M 87 is the first one discovered at optical frequencies. Both
 sources were among a few radio-loud AGNs known to posses the large scale X-ray
 jets before {\it Chandra} was launched. During the last decades, M 87 and
 Centaurus A were also being frequently observed at $\gamma$-rays, from sub-MeV
 to TeV photon energies, what resulted in positive detections or constraints on
 the flux upper limits.

\subsection{Centaurus A}

Centaurus A galaxy is a giant elliptical, which most probably mergered in the
 past with an other small and gas rich galaxy. With the distance $D = 3.5$ Mpc
 Centaurus A hosts the closest AGN, which is however difficult to be observed at
 IR and optical frequencies because of the obscuring dust lane. VLA studies
 \citep[e.g.,][]{bur83,cla92} show a double-lobed morphology of the Centaurus A
 radio source, with the jet and the counterjet of the FR I type and a kinetic
 power $\sim 2 \cdot 10^{43}$ erg/s. VLBI observations \citep[e.g.,][]{jon96}
 followed the jet and the counterjet to the pc-scales, suggesting the jet
 viewing angle $\theta \sim 50^0 - 80^0$. Optical and IR observations of the
 Centaurus A nucleus
 \citep[e.g.,][]{bai86,haw93,pac96,sch96,sch98,mar00,cap00} revealed the
 unresolved ($< 1$ pc) and variable central source with a small ($\sim 40$ pc)
 nuclear disc and a possible extremely compact nuclear torus. No IR/optical
 emission connected directly with the jet non-thermal radiation was found
 \citep[but see][]{joy91,sch96,sch98,mar00}. At the X-rays, the jet was observed
 in addition to the variable (at the time scales from minutes up to years)
 nuclear component \citep[respectively]{sch79,mor89}. Recently, {\it Chandra}
 telescope revealed details of the jet X-ray emission and presence of the X-ray
 counterjet \citep{kra02,hrd03}, as well as of the whole Centaurus A system
 \citep{kar02}. At higher energies, Centaurus A was observed by {\it CGRO}
 satellite, what allowed to detect the maximum of its nuclear $\gamma$-ray
 emission near $\sim 0.1$ MeV \citep{ste98}. For $\varepsilon_{\gamma} > 0.1$
 MeV the $\gamma$-ray flux decreases, although at $\sim 100$ MeV photon energies
 it is still detectable \citep{sre99}. {\it CANGAROO} observations of the
 extended ($\sim 14$ kpc) region around Centaurus A nucleus put the upper limit
 on the emission at the VHE range, $S(\varepsilon_{\gamma} \geq 1.5 \, {\rm
 TeV}) < 1.28 \cdot 10^{-11} \, {\rm ph \, cm^{-2} \, s^{-1}}$ \citep{row99}.
 However, the discussed object was detected in the past at TeV energies, with
 the observed flux $S(\varepsilon_{\gamma} \geq 0.3 \, {\rm TeV}) \sim 4.4 \cdot
 10^{-11} \, {\rm ph \, cm^{-2} \, s^{-1}}$ \citep{gri75}, what made Centaurus A
 the very first (although not confirmed) detected extragalactic source of the
 VHE radiation. Recent upper limit suggests that this emission can be variable
 on a timescale of years, in analogy to the low- and high-states of activity
 known from the TeV blazar observations \citep[but see also][for the X-ray
 variability of the large scale jet emission in M 87]{har97,har03}.

Existence of a hidden BL Lac core in the Centaurus A nucleus, as expected in a
 framework of unification scheme, was widely discussed over the last decade
 \citep{bai86,mor91,haw93,pac96,ste98,cap00}. Most recently, \citet{chi01}
 reconstructed broad-band spectrum of Centaurus A nucleus, from radio to
 $\gamma$-ray frequencies, and found spectacular similarities to the
 characteristic double-peaked blazar spectral energy distribution. The
 synchrotron component of this radiation was found to peak at far infra-red
 energy range, $\nu_{bl, \, br} \sim 10^{12} - 10^{13}$ Hz, with the observed
 luminosity $[\nu L_{\nu}]_{bl, \, br} \sim 10^{41} - 10^{42}$ erg/s. The
 inverse-Compton break energy was placed near $\sim 0.1$ MeV, with the observed
 power comparable to the synchrotron one. \citeauthor{chi01} fitted the SSC
 model to the multiwavelength Centaurus A nucleus emission, and found that all
 (except one) intrinsic parameters are similar to those of the low-luminous
 blazar sources. The only difference, as compared to the `typical' BL Lac
 broad-band spectrum, was a small value of the required Doppler factor,
 $\delta_{bl} \sim 1.6$. \citeauthor{chi01} interpreted their results as the
 evidence for the jet radial velocity structure at pc-scales, consisting of a
 fast central spine surrounded by a slower boundary layer \citep[see
 also][]{chi00}. Assuming that the physical properties (i.e. electron energy
 distribution, magnetic field intensity, etc) are the same in both jet
 components, the observed multiwavelength spectrum of the Centaurus A nucleus
 can then be regarded as a representation of a typical low-luminous blazar
 emission, but originating within the slower jet boundary layer and therefore
 less beamed as compared to the `classical' BL Lacs. It is consistent with the
 jet inclination $\sim 70^0$. Most probably, Centaurus A observed at small
 angles to the jet axis would be therefore classified as LBL, with the observed
 luminosity $\sim 10^{45} - 10^{46}$ erg/s and with the observed break frequency
 $\sim 10^{13} - 10^{14}$ Hz. Note, that in such a case the Centaurus A nucleus
 is not expected to radiate at the VHE range \citep[cf.][]{bai01}. For the
 estimates below, we assume $[\nu L_{\nu}]_{bl, \, 42} \sim 0.3$, $\nu_{bl, \,
 14} \sim 0.01$, $\Gamma_{bl} \sim 10$ and $\delta_{bl} \sim 1.6$.

In order to discuss the $\gamma$-ray emission of the large scale jet in
 Centaurus A, let us consider its brightest part in X-rays and at radio
 frequencies, the ensemble of the knots A1 - A4. The X-ray observations
 \citep{kra02} suggest that for this region $R_{-1} \sim 0.8$, $r_1 \sim 0.4$,
 $B_{-4} \sim 0.6$, $\alpha_X \sim 1.5$ and $L_X \sim 4 \cdot 10^{39}$ erg/s.
 The synchrotron break frequency is unknown, as there is no observation of the
 jet synchrotron emission at IR/optical frequencies. However, the existing upper
 limits \citep{sch96} are consistent with the synchrotron spectrum resulting
 from the electron energy distribution given by equation 1, with $\alpha_{R}
 \sim 0.5$, $\Delta \alpha \sim 1$ and the assumed $\nu_{syn, \, 14} \sim 1$.
 Thus, below we put $[\nu L_{\nu}]_{syn, \, br} / L_X \sim (10^{14} \, {\rm Hz}/
 2.4 \cdot 10^{17} \, {\rm Hz})^{1 - \alpha_X}$, implying for the A1 - A4 region
 $[\nu L_{\nu}]_{syn, \, 42} \sim 0.2$. Below we also take the galactic photons
 energy densities $U_{star}$ and $U_{dust}$ as estimated in section
 2.2.3\footnote{12 $\mu$m flux from Centaurus A is 23 Jy \citep{kna90}.}.

For the considered parameters of Centaurus A, the IC fluxes and break energies
 in function of the jet viewing angle $\theta$ are presented on Figure 1, for
 two different jet bulk Lorentz factors $\Gamma = 2, \, 6$. For small values of
 $\theta$ the EC(star) and EC(dust) emissions dominate, while for large jet
 inclinations the SSC and the EC(bl) processes are the more important ones. In
 general, small jet viewing angles are excluded as they imply non-observed large
 VHE flux due to comptonisation of the starlight. However, for large jet
 inclinations the SSC radiation also seems to overproduce (comparing to the
 present upper limits) the VHE photons. For example, for the usually considered
 $\theta = 70^0$ and $\Gamma \leq 2$, the expected SSC energy flux is $[\nu
 S_{\nu}]_{ssc, \, br} \sim 10^{-10} \, {\rm erg \, s^{-1} \, cm^{-2}}$ at the
 observed break energy $\varepsilon_{ssc, \, br} \sim 0.1 - 1$ TeV. This suggest
 presence of the KN effects decreasing the SSC and the EC(star) emissions, what
 in turn indirectly suggests relativistic jet velocities ($\Gamma > 2$, Figure
 2) and/or subequipartition magnetic field ($B \leq 10^{-5}$ G, Figure 3).
 Alternatively, the unobserved synchrotron break frequency $\nu_{syn, \, br}$ can
 be higher than assumed here $10^{14}$ Hz.

For the moderate and large jet inclinations, the EC(bl) radiation is expected to
 be very strong, with the observed flux $[\nu S_{\nu}]_{ec(bl), \, br} \sim
 10^{-11} - 10^{-10} \, {\rm erg \, s^{-1} \, cm^{-2}}$ and
 $\varepsilon_{ec(bl), \, br} \sim 10$ GeV (for $\Gamma = 1 - 10$). This
 emission could be therefore easily observed in the future by the {\it GLAST}
 satellite. The EC(dust) emission is important only for $\delta \gg 1$, and
 otherwise is dominated by the EC(bl) radiation. Finally, comptonisation of the
 CMB radiation does not play a role in the $\gamma$-ray emission of Centaurus A
 unless $\delta \gg 1$ or $B \ll B_{eq}$ (Figures 1 and 3).

\subsection{M 87}

M 87 galaxy is a giant elliptical, placed in the center of the Virgo Cluster
 X-ray emission. With the distance $D = 16$ Mpc it hosts one of the closest
 AGNs, with spectacular and widely studied large scale structure. VLA
 observations of the M 87 radio source revealed one-sided, highly polarised and
 edge-brightened radio jet \citep[e.g.,][]{owe89}, in addition to the large
 scale cavities inflated in the intergalactic medium \citep{owe00}. A relatively
 high radio luminosity places the discussed object at the borderline between FR
 I and FR II radio sources, although the one-sided jet with a kinetic power
 $\sim 5 \cdot 10^{44}$ erg/s displays FR I morphology. Detailed VLA and HST
 studies \citep[respectively]{bir95,bir99} revealed apparent superluminal
 motions of the jet components at distances up to a few hundreds parsecs from
 the active center, while VLBI observations measured much lower velocities at
 parsec scales. This suggests a complex velocity structure of the M 87 jet, with
 highly relativistic bulk velocities over all of its length ($\Gamma \sim$ a
 few) and a relatively small jet inclination to the line of sight \citep[$\theta
 \sim 30^0$; e.g.,][]{bic96}. IR observations of the considered object
 \citep{why01,cor02} give no evidence for the presence of an obscuring matter
 around the active center as expected for AGNs, consistently with a lack of the
 dusty nuclear torus or, eventually, with its small size ($< 50$ pc). Optical
 observations of the M 87 nuclear region revealed a small disc ($\sim 100$ pc)
 of ionised gas with a LINER emission line spectrum, fuelling a supermassive
 central black hole \citep[e.g.,][]{mac97}. The optical nuclear emission
 presents a smooth featureless continuum variable at time scales of months and
 connected with an unresolved ($\leq 5$ pc) central source \citep{tsv98}. The
 large scale jet is very prominent and highly polarised in optical, what allows
 to perform its detailed radio-to-optical spectroscopy and polarimetry
 \citep[e.g.,][]{mei96,spa96,per99,per01}. At X-rays the jet and the nucleus
 were early observed by {\it ROSAT} satellite \citep[e.g.,][]{sch82}. These
 observations allowed to construct broad-band radio-to-X-ray spectrum of the
 large scale jet emission \citep[e.g.,][]{bir91}, to put constraints on the
 X-ray nuclear radiation and its variability \citep{har97} and, finally, to
 study interaction between the radio-emitting matter and the surrounding thermal
 gas within the extended M 87 halo \citep{boh95,boh01}. Recently, the X-ray jet
 radiation was studied in more detail by {\it Chandra}
 \citep{mar02,wil02,har03}. Besides several attempts in the past, till the last
 year M 87 was not detected in $\gamma$-rays, neither by {\it CGRO} satellite
 \citep{sre96} nor by ground-based detectors in the VHE range. Up to now, the
 existing upper limit on the TeV emission of the discussed object was
 $S(\varepsilon_{\gamma} \geq 0.21 \, {\rm TeV}) < 1.2 \cdot 10^{-10} \, {\rm ph
 \, cm^{-2} \, s^{-1}}$ \citep{wee72}. However, recent {\it HEGRA} observations
 resulted in positive detection of the VHE emission from the M 87 radio galaxy,
 with the observed flux $S(\varepsilon_{\gamma} \geq 0.73 \, {\rm TeV}) \sim
 0.96 \cdot 10^{-12} \, {\rm ph \, cm^{-2} \, s^{-1}}$ \citep{aha03}.

Similarly to the case of Centaurus A, presence of a hidden blazar in the nucleus
 of M 87 was discussed previously by, e.g., \citet{tsv98,rey99,why01,cor02}.
 However, contrary to Centaurus A, a broad-band spectrum of the discussed
 central source in the IR-to-UV frequencies -- and therefore parameters of the
 considered hidden blazar -- is poorly constrained. One may note, that the
 radio-to-X-ray nuclear radiation is roughly similar to the large scale jet
 synchrotron emission, with $\alpha_{RO} < 1$ and $\alpha_{OX} > 1$
 \citep{tsv98,boh01}. This indicates, that the nuclear radio-to-X-ray continuum
 possibly results from the synchrotron emission of one electron population, as
 expected for the low-luminous blazars, with the observed synchrotron spectral
 break $\nu_{bl, \, br}$ placed near optical frequencies. The observed
 synchrotron power is then $ \sim 3 \cdot 10^{42}$ erg/s \citep{tsv98}.
 Furthermore, one may assume that, in analogy to Centaurus A, the observed M 87
 blazar-like emission originates from a slower component of the blazar jet and,
 hence, that the Doppler factor of the discussed radiation is close to unity.
 Here we take $\nu_{bl, \, 14} \sim 10$ and $[\nu L_{\nu}]_{bl, \, 42} \sim 1$
 with the assumed $\Gamma_{bl} \sim 10$ and $\delta_{bl} \sim 2$. The above
 values suggest that the considered M 87 blazar is situated on the borderline
 between LBL and HBL subclasses, and only large jet inclination preclude us to
 observe its `classical' blazar properties \citep[cf.][]{bai01}.

In order to evaluate the $\gamma$-ray flux of the M 87 large scale jet, let us
 consider its brightest knot A, dominating radio-to-X-ray radiative jet output.
 Besides differences in the jet morphology as observed in different frequencies,
 multiwavelength studies \citep[e.g.,][]{bir91,mei96,per01,mar02,wil02} allow
 one to construct broad-band spectrum of the considered region. They indicate
 $R_{-1} \sim 0.8$ (at radio and optical; X-ray knots seem to be twice
 narrower), $r_1 \sim 1$, $B_{-4} \sim 2.5$ and the synchrotron power $ \sim
 10^{42}$ erg/s. The exact value of the spectral break (and hence of the
 synchrotron break luminosity) depends on the detailed model of the synchrotron
 emission assumed for the knot regions \citep{per01,mar02}. However, the average
 optical spectral index $\alpha_O \sim 0.9$ \citep{per01} indicates that
 $\nu_{syn, \, br}$ is higher than $10^{14}$ Hz. Below, for a rough estimate, we
 put $\nu_{syn, \, 14} \sim 10$ and $[\nu L_{\nu}]_{syn, \, 42} \sim 0.3$
 \citep{wil02}. We also take the galactic photons energy densities $U_{star}$
 and $U_{dust}$ as estimated in section 2.2.3\footnote{12 $\mu$m flux from M 87
 is 0.42 Jy \citep{kna90}.}.

A similar analysis as for the Centaurus A $\gamma$-ray emission suggests, that
 in the case of M 87 large scale jet the EC(star) and EC(dust) processes should
 dominate production of the VHE photons (Figure 4). If the TeV flux detected
 recently by \citet{aha03} is in fact due to the knot A inverse-Compton
 emission, its most likely origin is comptonisation of the galactic stellar and
 circumstellar dust radiation. However, because of the expected KN effects, a
 detailed numerical analysis is required in order to find constraints on the jet
 bulk Lorentz factor and the magnetic field (cf. Figures 5 and 6). Possibilities
 that $\delta > 1$ and $B < B_{eq}$ could be additionally verified by the
 eventual detection of the EC(CMB) emission (Figures 4 and 6). The other IC
 processes for the M 87 jet are expected to be negligible because of the KN
 effects.

\section{Discussion and Conclusions}

In this paper we discussed the high energy $\gamma$-ray emission of the FR I
 large scale jets. We used the X-ray observations compiled with data from the
 other spectral bands in order to reconstruct energy distribution of
 ultrarelativistic electrons present in the considered objects. Our
 phenomenological approach to recover electron spectrum from observations
 rather than from theories of particle acceleration is dictated by the fact,
 that such theories are still insufficiently developed to enable quantitative
 analysis. Next, we analyzed possible origin of the seed photons contributing
 to the inverse-Compton emission of the obtained electron spectrum, including
 nuclear jet radiation as well as ambient, stellar and circumstellar emission
 of the host galaxies, and discussed in detail spectral properties of the
 resulting $\gamma$-ray radiative output. The approach take into account effects
 connected with relativistic bulk motion of the emitting region, correcting
 earlier computations presented in the literature. We also clearly indicated
 limitation of the adopted approach due the Klein-Nishina regime effects.
 Other restrictions of the presented model are connected with only roughly
 evaluated galactic radiation fields and hardly known parameters of the hidden
 blazar radiation. Also, the effects of $\gamma$-ray absorption on CIB radiation
 are only shortly mentioned in this paper. However, even with all the
 aforementioned uncertainties, one can conclude that the detection of the high
 energy $\gamma$-ray emission by future ground-based and space telescopes from,
 at least, some of the FR I large scale jets is possible. Thus, the future
 observations will provide important constraints on the unknown jet parameters,
 like the magnetic field intensity and the jet Doppler factor, confirming or
 excluding possibilities of $B \ll B_{eq}$ and $\delta \neq 1$ discussed in
 the literature.

Our analysis provides constraints for $\gamma$-ray emission of the nearby FR I
 sources Centaurus A and M 87. For Centaurus A we
 predict measurable -- by future $\gamma$-ray missions -- fluxes at $10$ GeV and
 $0.1 - 1$ TeV photon energies due to comptonisation of the blazar radiation and
 the synchrotron self-Compton process, respectively. In the case of M 87 we show
 that recently detected VHE emission can result from comptonisation of the
 stellar and circumstellar infrared photons of the host galaxy.

Possibility that the \emph{large scale} jets -- and not the active centers --
 in these two FR I radio galaxies can dominate production of high energy
 $\gamma$-rays was not discussed previously in the literature. Some other authors,
 e.g. \citet{bai01}, \citet{pro03} or \citet{don03}, suggested and studied
 production of very high energy $\gamma$-rays in the nuclear regions of the
 considered sources by the \emph{small scale}, blazar-like jets. This constitutes
 the main difference with our work. Unfortunatelly, the angular resolution of
 the {\it IAC} systems will not allow for separation of the kiloparsec scale
 jet $\gamma$-ray radiation from the eventuall core component. However, the
 core vs. the kpc-scale jet origin of $\gamma$-rays can be justified by studies
 of variability time scales.

\acknowledgments

The present work was supported by Komitet Bada\'{n} Naukowych through the grant
 PBZ-KBN-054/P03/2001, by the Department of Energy contract to SLAC no.
 DE-AC3-76SF00515, and by NASA Chandra grants via SAO grant no. GO1-2113X.
 Authors acknowledge useful comments and suggestions by F.A. Aharonian and the
 anonymous referee.

\appendix
\section{The observed and intrinsic jet luminosities}

Let us consider an observer at rest in a frame $K$, and an emitting fluid
 comoving frame denoted by $K'$. Fluid velocity in the observer frame is $\beta
 c$, the appropriate Lorentz factor is $\Gamma$, and the Doppler factor for a
 given inclination angle $\theta \equiv \cos^{-1} \mu$ with respect to the jet
 axis is $\delta = [\Gamma \, (1 - \beta \, \mu)]^{-1}$. In $K$ frame the
 radiation intensity is $I_{\nu}$ and the energy flux spectral density is
 $S_{\nu} = \int I_{\nu} \, d\Omega$, where $\Omega = (\cos^{-1} \mu , \phi)$.
 The total energy flux of the emission is $S = \int S_{\nu} \, d \nu$.

\begin{itemize}
\item As discussed by \citet{lin85} and \citet{sik97}, for a continuous
 (stationary) jet one has
\begin{equation}
S_{\nu} = {\delta^2 \over D^2} \, \int j'_{\nu'} \, dV \quad ,
\end{equation}
where $j'_{\nu'}$ is an intrinsic emissivity of the fluid (i.e. the one measured
 in $K'$), $dV$ is the observed emitting volume element and $D$ is the distance
 to the source. The observed isotropic luminosity is $L = 4 \pi \, D^2 \, S$.
 Hence, with $\nu = \delta \, \nu'$ and $dV = dV' / \Gamma$, one obtains
\begin{equation}
L = 4 \pi \, {\delta^3 \over \Gamma} \, {\partial L' \over \partial \Omega'}
 \quad ,
\end{equation}
where $\partial L' / \partial \Omega' \equiv \int j' \, dV'$ is the fluid
 intrinsic radiative power emitted in a given direction $\Omega' = (\cos^{-1}
 \mu' , \phi')$, and $j' = \int j'_{\nu'} \, d\nu'$. The total emitted power as
 measured in $K'$ is then
\begin{equation}
L' = \oint {\partial L' \over \partial \Omega'} d \Omega' = {\Gamma \over
 \delta^3} \, L \quad ,
\end{equation}
where the last equality holds for the intrinsically isotropic emission.
\item For a moving single blob of emitting plasma (with the observed volume
 element transforming as $dV = \delta \, dV'$) the observed energy flux spectral
 density is
\begin{equation}
S_{\nu} = {\delta^3 \over D^2} \, \int j'_{\nu'} \, dV' \quad .
\end{equation}
Hence
\begin{equation}
L = 4 \pi \, \delta^4 \, {\partial L' \over \partial \Omega'} \quad ,
\end{equation}
and in the case of intrinsically isotropic emission one obtains
\begin{equation}
L' = \oint {\partial L' \over \partial \Omega'} d \Omega' = {1 \over \delta^4}
 \, L \quad .
\end{equation}
\end{itemize}

Equations A3 and A6 specify the function $g_{cj / mb}(\Gamma, \theta)$ defined
 in section 2.2 for the two considered jet models. Note, that in both cases, the
 ratio between intrinsic power and comoving volume of the emitting region is
 $(L' / V') = (L / V) \, \delta^{-3}$. Hence, the equipartition magnetic field
 is
\begin{equation}
B_{eq} \propto \left({L'_{syn} \over (\nu'_{eq, \, max})^{1/2} \, V'}
 \right)^{2/7} = \left({L_{syn} \over (\nu_{eq, \, max})^{1/2} \, V}
 \right)^{2/7} \, \delta^{-5/7} \quad ,
\end{equation}
where $\nu'_{eq, \, max} = \nu_{eq, \, max} / \delta$ is the maximum synchrotron
 frequency considered in derivation of $B_{eq}$. As a result, the equipartition
 magnetic field measured in the emitting region rest frame is related to the
 equipartition value computed for no beaming by relation
\begin{equation}
B_{eq} = B_{eq, \, \delta = 1} \, \delta^{-5/7} \quad .
\end{equation}
This equation corrects the formula A7 of \citet{hk02}.

\section{External radiation fields}

Following the previous section, by $K'$ we denote the inertial rest frame of the
 emitting matter of the kpc-scale jet. In the observer rest frame, $K$, the jet
 has bulk velocity $\beta c$ and the respective bulk Lorentz factor $\Gamma$.
 Let us consider radiation external to the jet. Energy density of this emission
 measured in $K'$ is
\begin{equation}
U'_{ext} = {1 \over c} \, \int_{\Delta \Omega'} I'_{ext} \, d \Omega' \quad ,
\end{equation}
where $I'$ is the jet comoving intensity of the external radiation and the
 integration is performed over the photons arrival directions. Defining next
\begin{equation}
\delta_{in} = \left[ \Gamma \left( 1 - \beta \, \cos \psi_{in} \right)
 \right]^{-1} \quad ,
\end{equation}
where $\psi_{in}$ is the angle between external photons and the jet axis, one
 has $I'_{ext} = I_{ext} \, \delta_{in}^{-4}$, $d \Omega' = d \Omega \,
 \delta_{in}^2$ and thus
\begin{equation}
U'_{ext} = {1 \over c} \, \int_{\Delta \Omega'} I_{ext} \, \delta_{in}^{-2} \, d
 \Omega \quad ,
\end{equation}

\begin{itemize}
\item For emission isotropic in $K$ frame in vicinity of the kpc-scale jet, one
 obtains
\begin{equation}
U'_{ext} = {I_{ext} \over c} \, \int_{4 \pi} \delta_{in}^{-2} \, d \Omega =
 U_{ext} \, \Gamma^2 \, (1 + {\beta^2 \over 3}) \quad ,
\end{equation}
where the last equality is due to $U_{ext} = 4 \pi \, I_{ext} / c$.
\item In a case of the external emission illuminating the jet from its base one
 has $\psi_{in} \sim 0$ and hence
\begin{equation}
U'_{ext} = {1 \over c} \, \int_{\Delta \Omega'} { I_{ext} \, d \Omega \over (2
 \Gamma)^2} \approx {I_{ext} \, \Delta \Omega \over c} \, {1 \over (2 \Gamma)^2}
 \quad ,
\end{equation}
where $I_{ext} \, \Delta \Omega$ is approximately the emission energy flux
 received by the stationary observer located at the jet axis at the distance $r$
 from the considered source of the emission. The isotropic luminosity computed
 by such observer is therefore $L_{ext}(0) = 4 \pi \, r^2 \, I_{ext} \, \Delta
 \Omega$ and
\begin{equation}
U'_{ext} = {L_{ext}(0) \over 4 \pi \, r^2 \, c} \, {1 \over (2 \Gamma)^2} \quad
 .
\end{equation}
\end{itemize}

\section{Starlight emission}

Let us assume for the considered elliptical galaxies the spherical King-type
 stellar emissivity profile
\begin{equation}
j_{star}(\zeta) \propto \, \left(1 + \zeta^2\right)^{-3/2} \quad ,
\end{equation}
where $\zeta \equiv r / r_c$, $r \in (0, \, r_t)$ is a distance from the
 galactic center, $r_c$ is a core radius and $r_t$ is a terminal boundary radius
 of the stellar distribution (see section 2.2.3). With the assumed emissivity
 one can find the appropriate radiative intensity, $I_{star}(\zeta, \,
 \zeta_t)$, and next the mean intensity of the stellar radiation as a functions
 of $r$
\begin{equation}
J_{star}(\zeta, \, \zeta_t) = {1 \over 4 \pi} \, \int I_{star}(\zeta, \,
 \zeta_t) \, d\Omega \quad ,
\end{equation}
where $\zeta_t \equiv r_t / r_c$. The analytical computations were performed by
 \citet[see equation 26 therein]{tsa95}. The energy density of the considered
 emission is simply $U_{star}(\zeta, \, \zeta_t) = (4 \pi / c) \,
 J_{star}(\zeta, \, \zeta_t)$. Thus, knowing the central energy density of the
 starlight radiation, $U_{star, \, C}$, one can evaluate $U_{star}$ at any
 distance from the galactic center as
\begin{equation}
U_{star}(\zeta, \, \zeta_t) = U_{star, \, C} \, {J_{star}(\zeta, \, \zeta_t)
 \over J_{star}(0, \, \zeta_t)} \quad .
\end{equation}
For ellipticals with the bolometric luminosities ranging from $L_{bol} = 1.14
 \cdot 10^{44}$ erg/s to $L_{bol} = 1.23 \cdot 10^{45}$ erg/s, \citet{tsa95}
 estimated $U_{star, \, C} = 1.2 \cdot 10^{-7} - 5.9 \cdot 10^{-9}$ erg/cm$^3$,
 $r_c = 0.04 - 0.77$ kpc and $r_t = 44.4 - 123$ kpc, respectively (Table 1
 therein). For these parameters, at the distance $r = 1$ kpc from the galactic
 center, the bolometric starlight energy density computed accordingly to the
 equation C6 ranges from $0.9 \cdot 10^{-9}$ erg/cm$^3$ up to $3 \cdot 10^{-9}$
 erg/cm$^3$. Therefore, in this paper we take the average value $U_{star} (1 \,
 {\rm kpc}) \sim 10^{-9}$ erg/cm$^3$. We also assume for simplicity an
 approximately isotropic distribution of the discussed stellar emission in the
 galaxy rest frame at the kiloparsec scale.

\section{Anisotropy of the IC emission}

Let us consider the above defined rest frames $K$ and $K'$. The emitting fluid
 radiates through synchrotron and inverse-Compton processes, with intrinsic
 luminosities $[\nu' L'_{\nu'}]_{syn}$ and $[\nu' L'_{\nu'}]_{ic}$,
 respectively. The jet magnetic field energy density is $U'_B$, and energy
 density of the seed photons as measured in $K'$ is $U'_{seed}$. The ratio of
 the observed isotropic luminosities due to both considered processes is
\begin{equation}
{[\nu L_{\nu}]_{ic} \over [\nu L_{\nu}]_{syn}} = { 4 \pi \, g_{cj / mb}(\Gamma,
 \theta) \, { \partial [\nu' L'_{\nu'}]_{ic} \over \partial \Omega'} \over 4 \pi
 \, g_{cj / mb}(\Gamma, \theta) \, { \partial [\nu' L'_{\nu'}]_{syn} \over
 \partial \Omega' }} = { { \partial [\nu' L'_{\nu'}]_{ic} \over \partial \Omega'
 } \over { \partial [\nu' L'_{\nu'}]_{syn} \over \partial \Omega' }}
\end{equation}
(Appendix A). Note, that in the above equation one compares the synchrotron and
 the inverse-Compton luminosities due to the electrons with the same energy
 $\gamma \propto (\nu'_{syn} / B)^{1/2}$ and $\gamma \propto (\nu'_{ic} /
 \nu'_{seed})^{1/2}$ (for example $\gamma_{br}$ if break luminosities are
 considered). The intrinsic synchrotron power emitted in a given direction
 $\Omega' \equiv (\cos^{-1} \mu', \phi')$ by isotropic electrons with the energy
 distribution $n'_e(\gamma)$ within an uniformly filled volume $V'$ is
\begin{equation}
{\partial [\nu' L'_{\nu'}]_{syn} \over \partial \Omega' } = V' \, [\nu'
 j'_{\nu'}]_{syn} = {1 \over 2} \, {[\gamma n'_e(\gamma)] \over 4 \pi} \, V' \,
 | \dot{\gamma} |_{syn} \, m c^2 \quad ,
\end{equation}
where $|\dot{\gamma}|_{syn} \, m c^2= {4 \over 3} c \sigma_T \, U'_B \,
 \gamma^2$ is a mean rate of electron energy losses due to the synchrotron
 emission. It is not difficult to show (see below), that in the Thomson regime
 considered in this paper an analogous expression representing comptonisation of
 anisotropic (in $K'$) seed photons is
\begin{equation}
{\partial [\nu' L'_{\nu'}]_{ic} \over \partial \Omega' } = V' \, [\nu'
 j'_{\nu'}(\Omega')]_{ic} = {1 \over 2} \, {[\gamma n'_e(\gamma)] \over 4 \pi}
 \, V' \, |\dot{\gamma}(\Omega')|_{ic} \, m c^2 \quad ,
\end{equation}
where $|\dot{\gamma}(\Omega')|_{ic}$ denotes electron cooling rate due to the
 inverse-Compton emission. For the seed photons antiparallel ($+$) or parallel
 ($-$) in $K'$ to the jet axis (if the seed photons are distributed
 isotropically around the relativistic jet, in its comoving frame their
 directions are almost antiparallel to the jet axis) one has
\begin{equation}
|\dot{\gamma}(\Omega')|_{ic(\pm)} = {c \sigma_T \over m c^2} \, U'_{seed} \,
 \gamma^2 \, (1 \pm \mu')^2 \quad ,
\end{equation}
while in the case of the isotropic ($iso$) distribution of the seed photons in
 the jet rest frame
\begin{equation}
|\dot{\gamma}(\Omega')|_{ic(iso)} = {4 \over 3} \, {c \sigma_T \over m c^2} \,
 U'_{seed} \, \gamma^2 \quad.
\end{equation}
Thus, the function $f_{\pm / iso}(\Gamma, \theta)$ defined in equation 16 is
\begin{equation}
f_{\pm / iso}(\Gamma, \theta) = \left\{ \begin{array}{ccc} {3 \over 4} \, (1 \pm
 \mu')^2 & (\pm) \\ 1 & (iso) \end{array} \right. \quad .
\end{equation}
Transforming it to the observer rest frame, with $\mu' = (\mu - \beta) / (1 -
 \beta \mu)$ one finally obtains $f_{\pm / iso}(\Gamma, \theta)$ as given by the
 equation 17 \citep[see also][]{der92,der95}.

In order to find $|\dot{\gamma}(\Omega')|_{ic}$ for the cases discussed above,
 let us evaluate the appropriate inverse-Compton emissivity $[j'_{\nu'}]_{ic}$.
 Here we write $[\nu' j'_{\nu'}(\Omega')]_{ic} = m c^2 \, \epsilon'^2 \,
 \dot{n}'_{ic} (\epsilon', \Omega')$, where the dimensionless photon energy
 $\epsilon' \equiv h \nu' / m_e c^2$ and the photon emissivity is
\begin{eqnarray}
\dot{n}'_{ic} (\epsilon', \Omega') & = & c \, \int d \epsilon'_{seed} \, \oint d
 \Omega'_{seed} \, \int d \gamma \, \oint d \Omega'_{e}{} \nonumber\\ & & {}(1 -
 \cos \chi') \, \sigma \, n'_{seed} (\epsilon'_{seed}, \Omega'_{seed}) \, n'_{e}
 (\gamma, \Omega'_e) \quad .
\end{eqnarray}
Here $n'_{seed} (\epsilon'_{seed}, \Omega'_{seed})$ is a seed photon number
 density, $n'_e (\gamma, \Omega'_e)$ is the electron energy distribution,
 $\chi'$ is the angle between the electron and the seed photon direction,
 $\sigma$ is the inverse-Compton scattering cross-section, and all quantities
 are measured in the emitting fluid rest frame $K'$ \citep{der95}.

\begin{itemize}
\item For the seed photons antiparallel ($+$) or parallel ($-$) in $K'$ to the
 jet axis, assuming azimutal symmetry and characteristic seed photon energy
 $\langle \epsilon'_{seed} \rangle$, one can write
\begin{equation}
n'_{seed}(\epsilon'_{seed}, \Omega'_{seed}) = {1 \over 2 \pi} \, n'_{seed} \,
 \delta \left( \epsilon'_{seed} - \langle \epsilon'_{seed} \rangle \right) \,
 \delta \left(\mu'_{seed} \pm 1 \right) \quad ,
\end{equation}
where $\Omega'_{seed} = (\cos^{-1} \mu'_{seed}, \phi'_{seed})$. Thus, with the
 isotropic electron energy distribution $n'_{e}(\gamma, \Omega'_e) =
 n'_e(\gamma) / 4 \pi$ and the Thomson scattering cross-section
\begin{equation}
\sigma = \sigma_T \, \delta \left[ \epsilon' - \gamma^2 \, \epsilon'_{seed} \,
 (1 - \cos \chi') \right] \, \delta \left[ \Omega' - \Omega'_e \right]
\end{equation}
\citep[e.g.,][]{der95}, one obtains
\begin{eqnarray}
\dot{n}'_{ic} (\epsilon', \Omega') & = & {1 \over 2} \, {c \, \sigma_T \over 4
 \pi} \, n'_{seed} \, \left( { 1 \pm \mu' \over \epsilon' \, \langle
 \epsilon'_{seed} \rangle} \right)^{1/2}{} \nonumber\\ & & {}\int d \gamma \,
 n'_{e} (\gamma) \, \delta \left[ \gamma - \sqrt{{\epsilon' \over \langle
 \epsilon'_{seed} \rangle \, (1 \pm \mu')}} \right] \quad .
\end{eqnarray}
This relation, with $\gamma = \sqrt{\epsilon' / \langle \epsilon'_{seed} \rangle
 \, (1 \pm \mu')}$ and $U'_{seed} = n'_{seed} \, \langle \epsilon'_{seed}
 \rangle \, m c^2$, leads to $\partial [\nu' L'_{\nu'}]_{ic} / \partial \Omega'$
 given by expressions D3 and D4.
\item For the seed photons isotropic in the jet comoving frame ($iso$), one can
 write
\begin{equation}
n'_{seed}(\epsilon'_{seed}, \Omega'_{seed}) = {1 \over 4 \pi} \, n'_{seed} \,
 \delta \left( \epsilon'_{seed} - \langle \epsilon'_{seed} \rangle \right) \quad
 .
\end{equation}
Thus, with $n'_{e}(\gamma, \Omega'_e) = n'_e(\gamma) / 4 \pi$ and the Thomson
 scattering cross-section
\begin{equation}
\sigma = \sigma_T \, \delta \left[ \epsilon' - {4 \over 3} \, \gamma^2 \,
 \epsilon'_{seed}\right] \, \delta \left[ \Omega' - \Omega'_e \right] \quad ,
\end{equation}
one obtains
\begin{eqnarray}
\dot{n}'_{ic(iso)} (\epsilon', \Omega') & = & {1 \over 2} \, {c \, \sigma_T
 \over 4 \pi} \, n'_{seed} \, \left( { 3 / 4 \over \epsilon' \, \langle
 \epsilon'_{seed} \rangle} \right)^{1/2}{} \nonumber\\ & & {}\int d \gamma \,
 n'_{e} (\gamma) \, \delta \left[ \gamma - \sqrt{{\epsilon' \over {4 \over 3} \,
 \langle \epsilon'_{seed} \rangle}} \right] \quad .
\end{eqnarray}
This, with $\gamma = \sqrt{\epsilon' / {4 \over 3} \, \langle \epsilon'_{seed}
 \rangle}$ and $U'_{seed} = n'_{seed} \, \langle \epsilon'_{seed} \rangle \, m
 c^2$ leads to $\partial [\nu' L'_{\nu'}]_{ic} / \partial \Omega'$ given by
 expressions D3 and D5.
\end{itemize}

\section{Break frequencies of the IC radiation}

Let us consider the break frequency of the large scale jet inverse-Compton
 emission. Accordingly to the formalism presented in Appendix D, in the jet
 comoving frame a seed photon with the characteristic (break) frequency
 $\nu'_{seed}$ is upscattered by the ultrarelativistic electron with Lorentz
 factor $\gamma$ to the frequency $\gamma^2 \, \nu_{seed}' \, (1 - \cos \chi')$.
 For the electron break Lorentz factor $\gamma_{br}$ one obtains the observed
 break frequency of the inverse-Compton emission
\begin{equation}
\nu_{ic(seed), \, br} = \gamma_{br}^2 \, \nu'_{seed} \, \delta \, (1 - \cos
 \chi') \quad .
\end{equation}

\begin{itemize}
\item For comptonisation of radiation distributed isotropically around the jet,
 one has $\nu'_{seed} = \Gamma \, \nu_{seed}$ and $1 - \cos \chi' = 1 + \mu'$.
 Therefore
\begin{equation}
\nu_{ec(seed), \, br} = \gamma_{br}^2 \, \nu_{seed} \, \Gamma \, \delta \, (1 +
 \mu') = \gamma_{br}^2 \, \nu_{seed} \, \delta^2 \, {(1 + \mu) \over (1 +
 \beta)} \quad .
\end{equation}
\item In the case of seed photons illuminating kpc-scale jet from behind
 $\nu'_{seed} = \nu_{seed} / \Gamma$ and $1 - \cos \chi' = 1 - \mu'$. Hence
\begin{equation}
\nu_{ec(seed), \, br} = \gamma_{br}^2 \, \nu_{seed} \, {1 \over \Gamma} \,
 \delta \, (1 - \mu') = \, \gamma_{br}^2 \, \nu_{seed} \, \delta^2 \, (1 - \mu)
 \, (1 + \beta) \quad .
\end{equation}
\item For comptonisation of the isotropic (in $K'$) radiation one has
 $\nu'_{seed} = \nu_{seed} / \delta$ and the averaged $\langle \nu'_{ic(seed)} /
 \nu'_{seed} \rangle = (4 / 3) \, \gamma^2$. Therefore
\begin{equation}
\nu_{ic(seed), \, br} = {4 \over 3} \gamma_{br}^2 \, \nu_{seed} \quad .
\end{equation}
\end{itemize}

Equations E2 - E4 specify the function $h_{\pm / iso}(\Gamma, \theta)$ defined
 in section 2.3.

\begin{figure}
\plotone{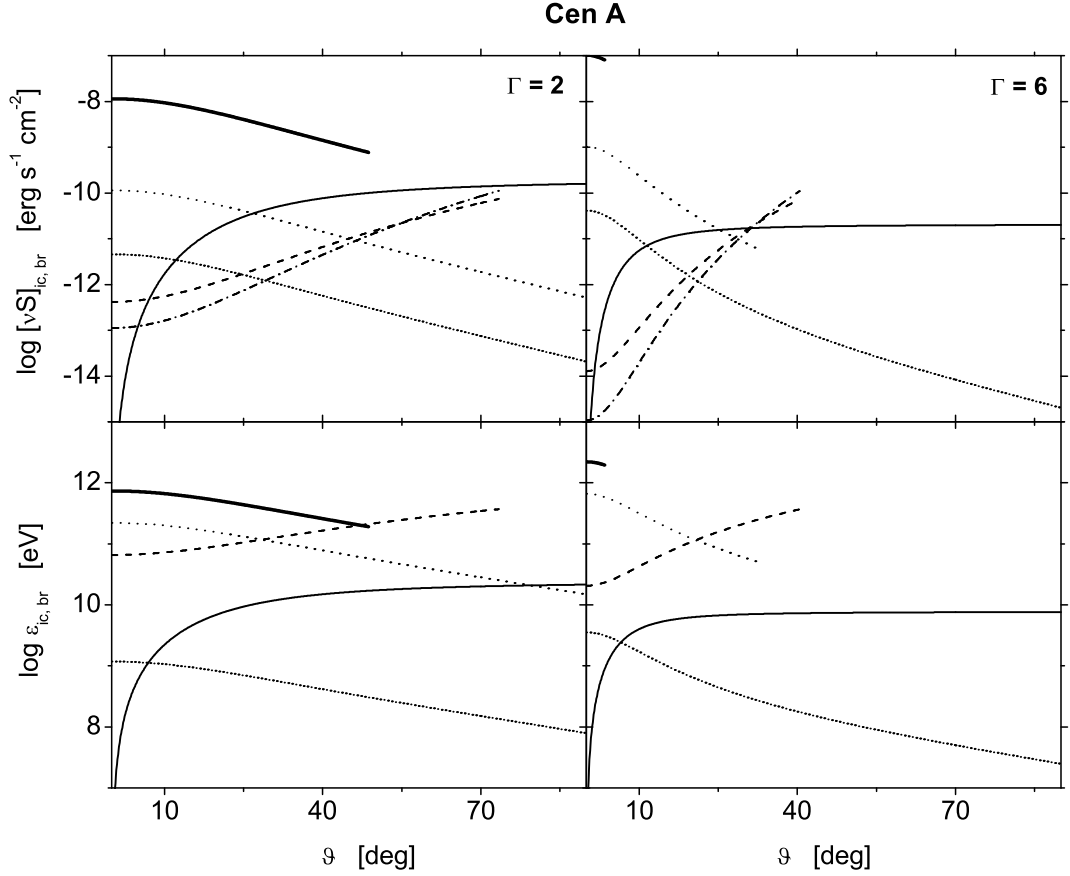}
\caption{ High energy $\gamma$-ray emission of the Centaurus A large scale jet
 region A1 - A4 for different jet viewing angles $\theta$, $B = B_{eq}$ and the
 bulk Lorentz factors $\Gamma = 2 \, 6$. Other parameters are discussed in
 Section 3.1. The upper panels show the observed IC fluxes, while the lower
 panels indicate the respective observed photon break energies. The solid lines
 correspond to the EC(bl) emission, the dotted lines to the EC(dust) process,
 the short-dotted line to the EC(CMB) radiation, thick solid line to the
 EC(star) emission and, finally, the dashed and dash-dotted lines to the SSC
 radiation of a continuous jet and moving blob, respectively. The lines are
 presented only till the Klein-Nishina limit. \label{fig1}}
\end{figure}

\begin{figure}
\plotone{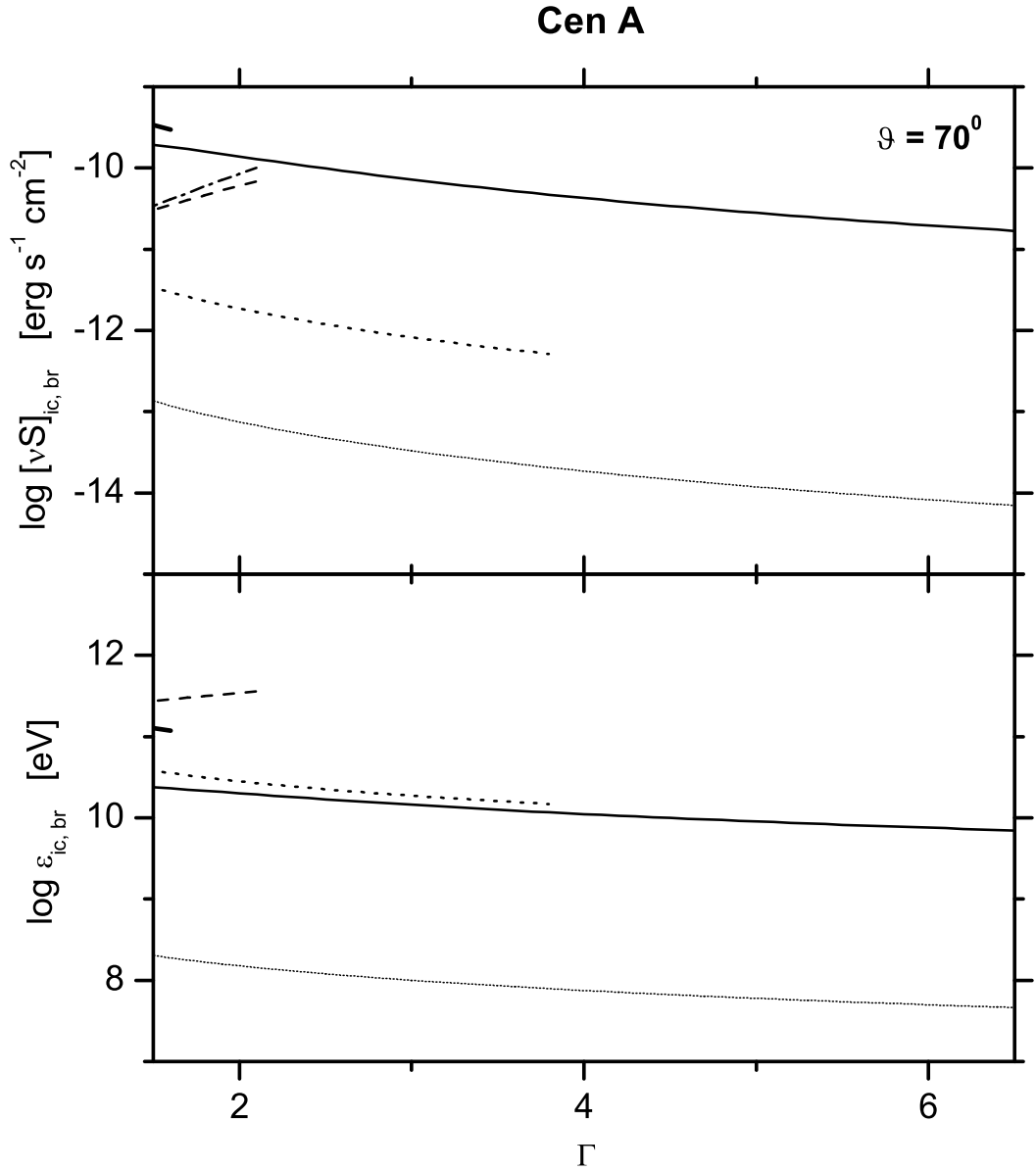}
\caption{ High energy $\gamma$-ray emission of the Centaurus A large scale jet
 region A1 - A4 in fuction of the bulk Lorentz factor $\Gamma$, for the jet
 viewing angle $\theta = 70^0$ and the magnetic field $B = B_{eq}$. The other
 parameters and description of the presented curves are the same as in Figure 1.
 \label{fig2}}
\end{figure}

\begin{figure}
\plotone{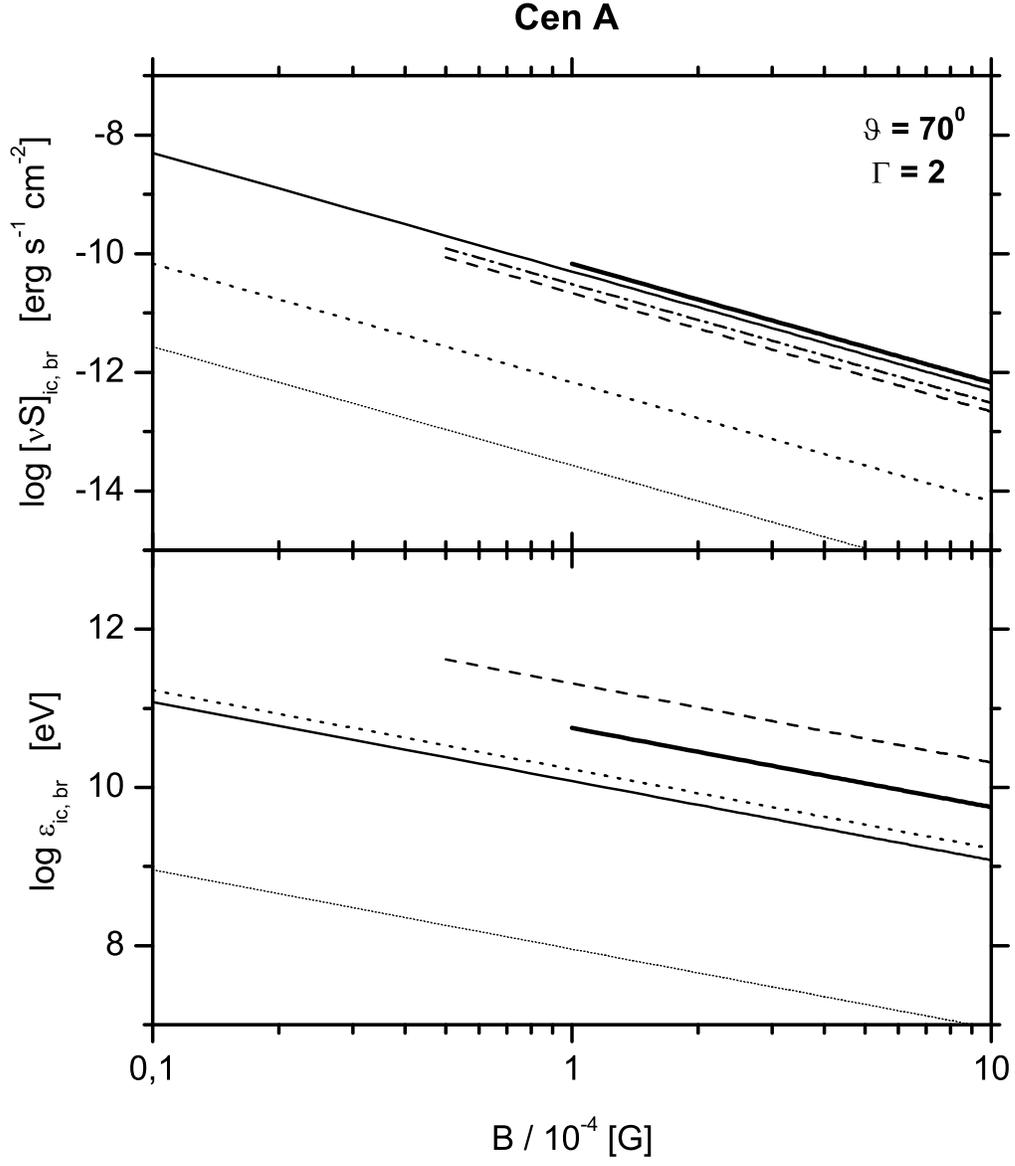}
\caption{ High energy $\gamma$-ray emission of the Centaurus A large scale jet
 region A1 - A4 in fuction of the magnetic field $B$, for the jet viewing angle
 $\theta = 70^0$ and the bulk Lorentz factor $\Gamma = 2$. The other parameters
 and description of the presented curves are the same as in Figure 1.
 \label{fig3}}
\end{figure}

\begin{figure}
\plotone{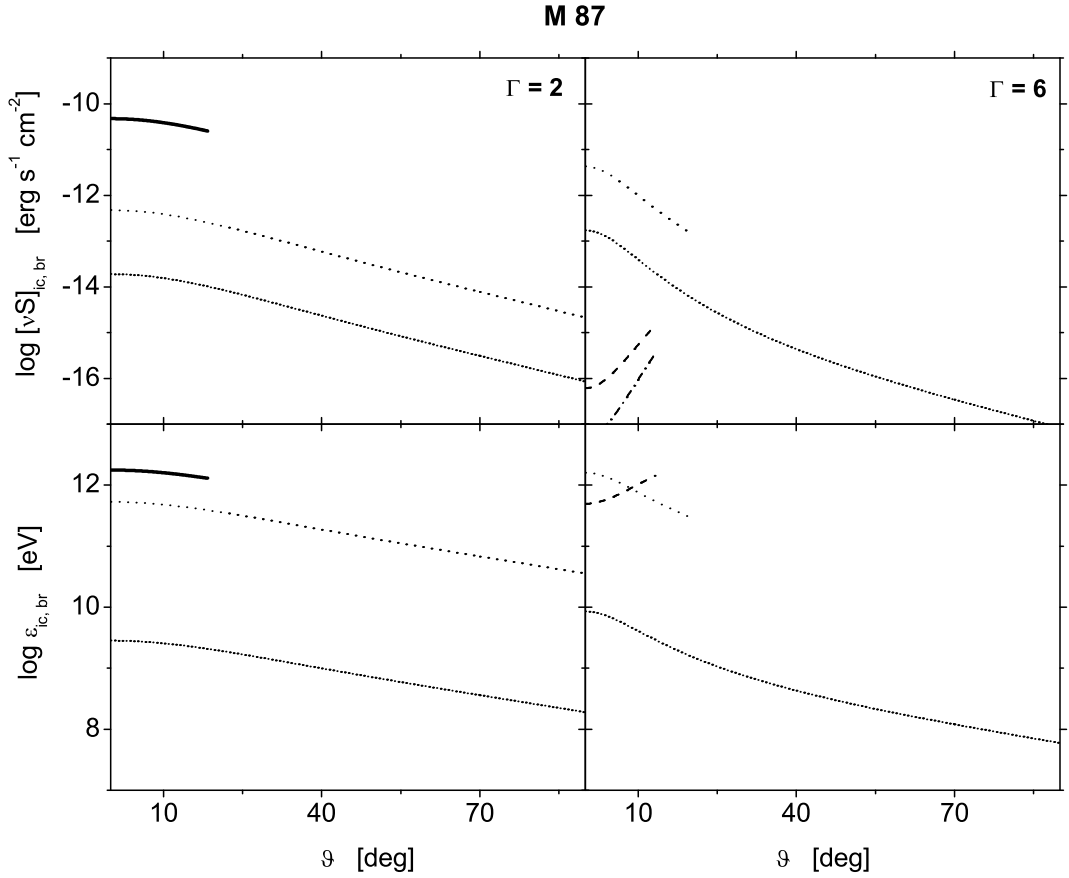}
\caption{ High energy $\gamma$-ray emission of the M 87 knot A for different jet
 viewing angles $\theta$, $B = B_{eq}$ and the bulk Lorentz factors $\Gamma = 2
 \, 6$. The other parameters are discussed in Section 3.2. Description of the
 presented curves is the same as in Figure 1. \label{fig4}}
\end{figure}

\begin{figure}
\plotone{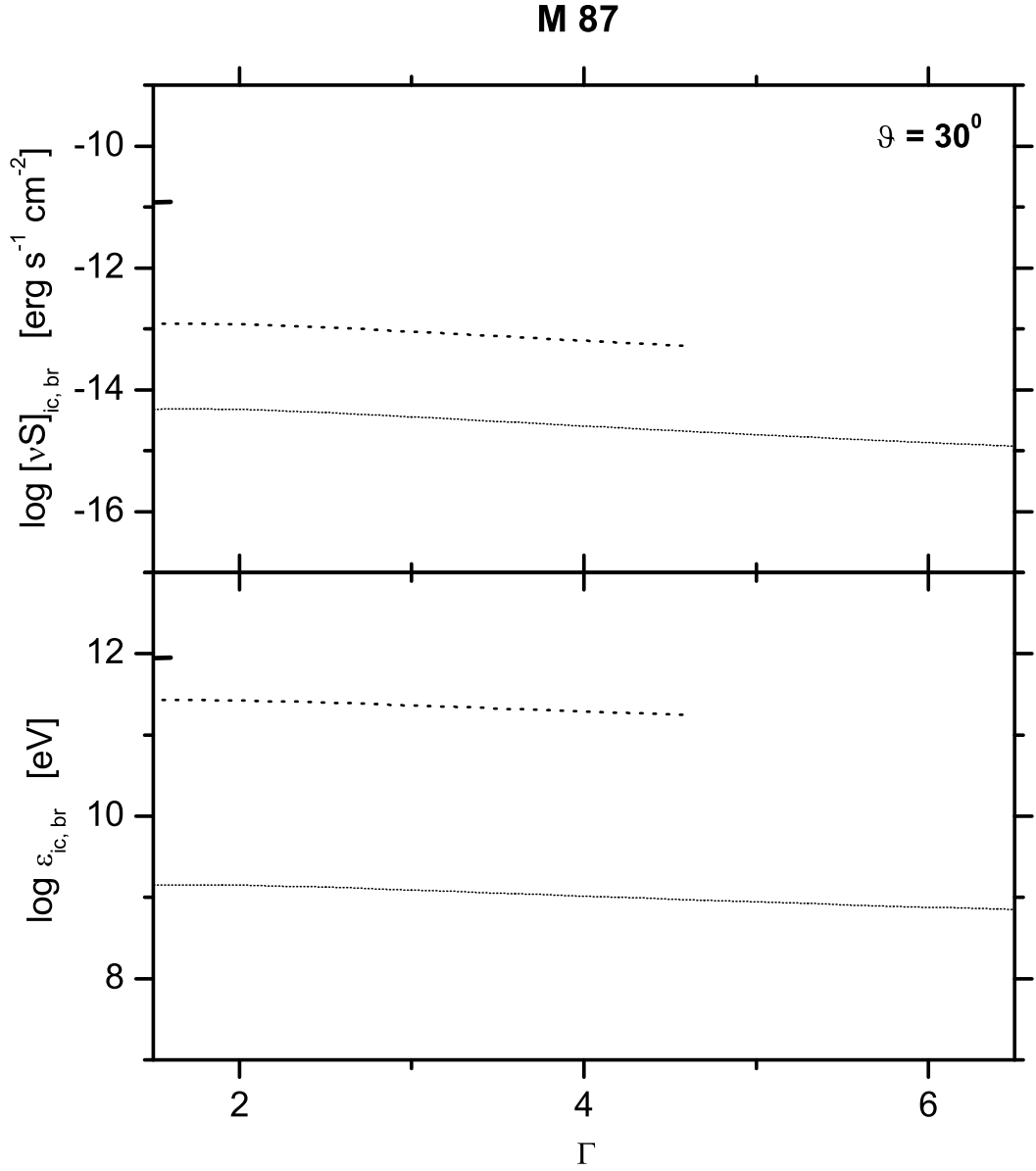}
\caption{ High energy $\gamma$-ray emission of the M 87 knot A in fuction of the
 bulk Lorentz factor $\Gamma$, for the jet viewing angle $\theta = 30^0$ and the
 magnetic field $B = B_{eq}$. The other parameters and description of the
 presented curves are the same as in Figure 4. \label{fig5}}
\end{figure}

\begin{figure}
\plotone{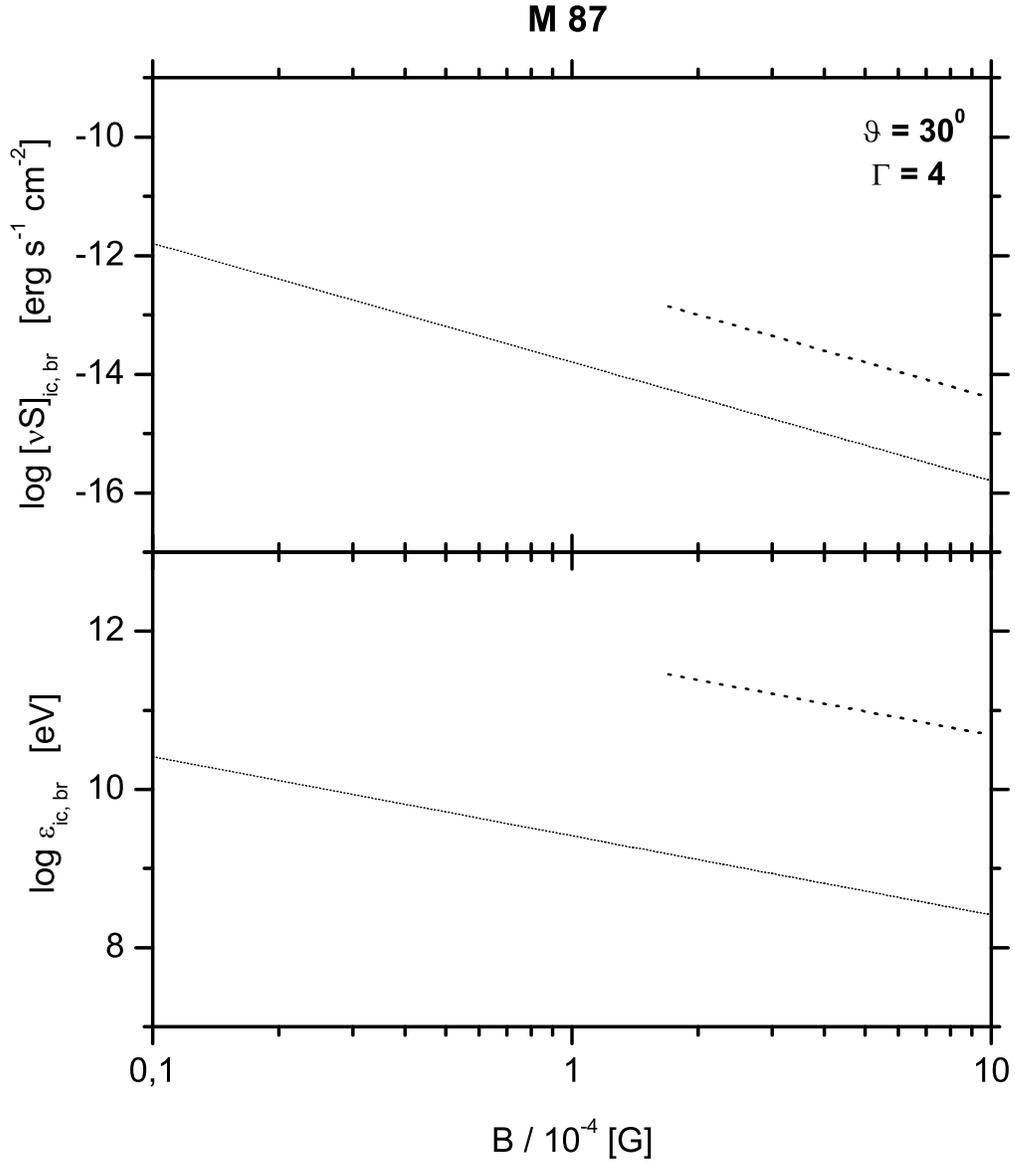}
\caption{ High energy $\gamma$-ray emission of the M 87 knot A in fuction of the
 magnetic field $B$, for the jet viewing angle $\theta = 30^0$ and the bulk
 Lorentz factor $\Gamma = 4$. The other parameters and description of the
 presented curves are the same as in Figure 4. \label{fig6}}
\end{figure}

\end{document}